\documentclass[a4paper, reprint, showkeys, nofootinbib, twoside,superscriptaddress]{revtex4-2}
\usepackage[english]{babel}
\usepackage[utf8]{inputenc}
\usepackage[colorinlistoftodos, color=green!40, prependcaption]{todonotes}
\usepackage[pdftex, pdftitle={Article}, pdfauthor={Author}]{hyperref} 
\usepackage{amsthm}
\usepackage{mathtools}
\usepackage{physics}

\usepackage{graphicx}
\usepackage[left=23mm,right=13mm,top=35mm,columnsep=15pt]{geometry} 
\usepackage{adjustbox}
\usepackage{placeins}
\usepackage{lipsum}
\usepackage{csquotes}
\usepackage{amsmath}
\usepackage{amsfonts, amssymb}
\usepackage{color}
\usepackage{float}
\usepackage{array}
\usepackage{colortbl}
\usepackage{verbatim}
\usepackage{setspace}
\usepackage{subfigure}
\usepackage{mathrsfs}

\definecolor{Cyan}{cmyk}{1,0,0,0}
\definecolor{SkyBlue}{cmyk}{0.62,0,0.12,0}
\definecolor{Salmon}{cmyk}{0,0.53,0.38,0}
\definecolor{Apricot}{cmyk}{0,0.32,0.52,0}
\definecolor{Lavender}{cmyk}{0,0.48,0,0}
\definecolor{CarnationPink}{cmyk}{0,0.63,0,0}
\definecolor{GreenYellow}{cmyk}{0.15,0,0.69,0}
\definecolor{LightCyan}{rgb}{0.88,1,0}
\definecolor{SpringGreen}{cmyk}{0.26,0,0.76,0}
\definecolor{PinkLace}{cmyk}{0, 0.204, 0.043, 0}
\definecolor{RasberryRed}{cmyk}{ 0,0.88,0.81,0.02}
\definecolor{Yellow}{cmyk}{ 0,0.1,1,0}
\definecolor{Green}{cmyk}{ 0.19,0,0.71,0.21}
\definecolor{Magenta}{cmyk}{ 0.12,0.60,0,0}
\definecolor{Blue}{cmyk}{ 0.75,0.75,0,0.20}

\newcommand{\scD}{\mathcal{D}}
\newcommand{\scA}{\mathcal{A}}
\newcommand{\scN}{\mathcal{N}}
\newcommand{\scC}{\mathcal{C}}
\newcommand{\mscC}{\mathscr{C}}
\newcommand{\mscN}{\mathscr{N}}

\usepackage[inline]{trackchanges}
\addeditor{EM}
\addeditor{LC}
\addeditor{LK}

\begin{document}
\title{Dielectrowetting of a thin nematic liquid crystal layer}

\author{E. Mema}
    \affiliation{United States Military Academy, West Point, NY}
\author{L. Kondic}
    \affiliation{Department of Mathematical Sciences, New Jersey Institute of Technology, Newark, NJ}
\author{L.J. Cummings}
    \affiliation{Department of Mathematical Sciences, New Jersey Institute of Technology, Newark, NJ}


\begin{abstract}
	We consider a mathematical model that describes the flow of a Nematic Liquid Crystal (NLC) film placed on a flat substrate, across which a spatially-varying electric potential is applied. Due to their polar nature, NLC molecules interact with the (nonuniform) electric field generated, leading to instability of a flat film.
	Implementation of the long wave scaling leads to a partial differential equation that predicts the subsequent time evolution of the thin film. This equation is coupled to a boundary value problem that describes the interaction between the local molecular orientation of the NLC (the director field) and the electric potential.  We investigate numerically the behavior of an initially flat film for a range of film heights and surface anchoring conditions.
\end{abstract}


\maketitle

\section{Introduction}

Dielectrowetting, a consequence of dielectrophoresis, involves the use of a nonuniform electric field to control spreading of dielectric fluids (such as nematic liquid crystals (NLCs), among others), on a substrate. A common focus of dielectrowetting experiments is the controlled manipulation of dielectric droplets using the interaction with an electric field to spread them into a thin film (see {\it e.g.} Brown {\it et al.}~\cite{Brown2009}); or vice versa:
to break down a large parent droplet into smaller droplets (see {\it e.g.} McHale {\it et al.} \cite{McHale2011}), with numerous possible applications~\cite{Cheng06,Ren2011,Ren2013,Russell2014,Russell2015,Heikenfeld2013}. The classic dielectrowetting experiment, described in pioneering works such as those of Cheng {\it et al.} \cite{Cheng06} and Brown {\it et al.} \cite{Brown2009}, involves a dielectric droplet or film spreading over a flat substrate, which contains  interdigitated electrodes. A potential difference applied across these electrodes generates a nonuniform electric field, which acts on the polar molecules, driving flow.  An excellent and comprehensive recent review of the field is provided by Edwards {\it et al.} \cite{Edwards2018}. A key feature of dielectrowetting that makes it desirable for applications is that it works with any dielectric liquid, in contrast to the more well-known ``Electrowetting on Dielectric'' (EWOD) technology, which relies on the movement of free charge and is limited to conducting liquids only. Dielectrowetting also offers the capability of spreading a partially-wetting droplet into a fully-wetting film (and vice-versa)~\cite{McHale2013}, which opens doors for new thin-film-based devices in the fields of microfluidics and optofluidics \cite{Edwards2018}.

Theoretical investigations into dielectrowetting to date focus mostly on isotropic dielectric liquids  \cite{Brown2011,Brown2015,McHale2011,McHale2013}. Considering their wide industrial use, it is imperative to develop mathematical models that describe the behavior of anisotropic dielectric liquids, such as NLCs, in a similar setting. NLCs show rather complex behavior compared to Newtonian fluids because they consist of rodlike molecules, which have a dipole moment. In the absence of an electric field the molecules tend to align locally, due to interactions between the dipoles, which imparts elasticity to the material. Molecular alignment is also mediated by surface effects; so-called \emph{surface anchoring}~\cite{DeGennes1995,stewart,Chandrasekhar}. 
Applying an electric field also affects molecular alignment: the long axis of the NLC molecules will align parallel or perpendicular to the local electric field direction (depending on the dipole moment). Application of an electric field near a bounding surface sets up a competition between the dielectric force on the NLC molecules and the surface anchoring at the boundary. As a result, mathematical models of NLCs in such settings can be complex.

In this paper, we seek to develop a mathematical model that describes the flow of a thin NLC film under the effect of a nonuniform electric field. We consider a setup similar to that of Brown {\it et al.} \cite{Brown2010,Brown2009} where a thin layer of dielectric liquid is placed on a substrate containing parallel interdigitated electrodes, which generate a periodic electric field profile. We make use of the commonly considered  long wave scaling to derive a version of the thin film equation governing the evolution of the film height.  This equation is coupled to a pair of boundary value problems, one for the electric potential within the film, and one describing the average orientation of the long axis of the NLC molecules (modeled in the theory by the {\it director field} $\mathbf{n}$, a unit vector). The modeling is simplified by exploiting a disparity in timescales: both the director field and electric field are assumed to be in instantaneous equilibrium for the current free surface geometry. The resulting model is still complex, and numerical techniques are used to explore how the initial film height and anchoring conditions at the boundaries influence the evolution of the free surface over time.

\section{Mathematical Model}
\label{mathmodel}

\begin{figure}[h!]
\centering
    \subfigure[~]{
	    \includegraphics[width=0.5\textwidth]{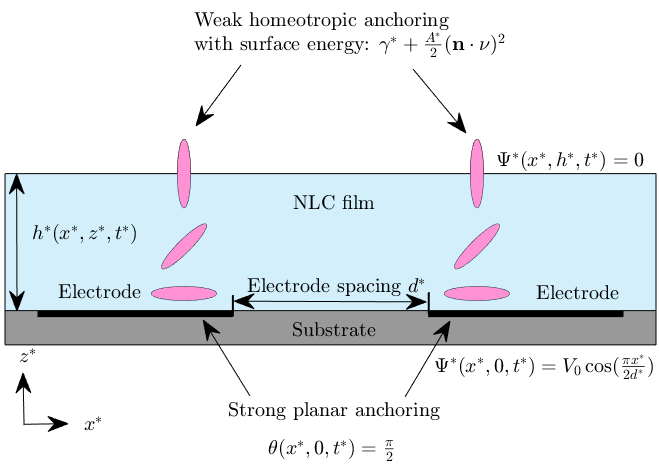}}
    \subfigure[~]{
        \includegraphics[width=0.35\textwidth]{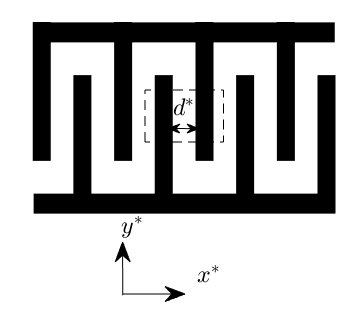}}
	\caption{(a)~Diagram showing model setup and key parameters in dimensional coordinates. (b)~Plan view of experimental setup showing the electrode geometry. Dashed rectangle indicates the region that corresponds to (a). 
		}
	\label{diagram}
\end{figure}

In what follows, we develop a mathematical model that describes the flow and free surface evolution of a nematic liquid crystal (NLC) film under the effect of a nonuniform electric field. The basic setup, shown in Figure~\ref{diagram}, consists of a thin NLC layer placed on a horizontal substrate. The substrate contains a pair of planar, interdigitated electrodes, leading to an electric potential that is nonuniform in the plane of the substrate.
In our model we will assume that the electrodes are of infinite extent (in the $y^*$ direction in Fig.~\ref{diagram}), so that the electric field generated by the applied potential will vary only in $x^*$, the coordinate in the plane of the substrate perpendicular to the electrodes, and $z^*$, which measures distance perpendicular to the substrate into the NLC. The main dependent variables that govern the dynamics are the velocity and director field of the NLC, together with the electric potential generated by the electrodes. In line with our assumed electrode geometry, throughout this paper we restrict attention to the 2D case in which the NLC velocity field $\mathbf{v^*} = (u^*,w^*)$ and the director field $\mathbf{n} = (\sin \theta,\cos \theta)$ are confined to the $(x^*,z^*)$-plane; so that the electric potential at time $t^*$ may be written as $\Psi^* (x^*,z^*,t^*)$; and the director orientation $\theta$, which represents the angle that the director makes with the $z^*$ axis, is $\theta(x^*,z^*,t^*)$. Here and below, $*$ superscripts are used to denote dimensional quantities. Quantities without superscripts are dimensionless. 

\subsection{Leslie-Ericksen Equations}

The flow of NLC may be described by the standard Leslie-Ericksen model \cite{Leslie,Ericksen1960}, which comprises the energy, (zero inertia) momentum balance and mass conservation equations as follows: 
\begin{align}
-\frac{\partial W^*}{\partial \bf{n}} + \mathbf{\nabla^*} \cdot \left( \frac{\partial W^*}{\partial \nabla^* \mathbf{n}}\right)+\mathbf{G^*} &= 0,\label{energy}\\
-\mathbf{\nabla^*} W^* + (\mathbf{\nabla^*\mathbf{n}})\cdot \mathbf{G}^* + \mathbf{\nabla^*}\cdot\mathbf{\sigma^*} &= 0,\label{momentum}\\
\mathbf{\nabla^*}\cdot \mathbf{v^*} &= 0,\label{mass}
\end{align}
where $W^*$ is the bulk free energy density and $\sigma^*$ is the stress tensor; these and the other quantities in Eqs.~(\ref{energy})--(\ref{mass}) are defined below:
\begin{eqnarray}
	&W^* = W^*_{\rm e} + W^*_{\rm d},\label{energy_density}\\
	&2 W^*_{\rm e} = K^*[(\nabla^* \cdot \mathbf{n})^2 +  (\mathbf{n}\times(\nabla^*\times \mathbf{n}))^2],
	 \label{elastic_density}\\
	&2 W_{\rm d}^*= -\varepsilon_0^* (\varepsilon_{\parallel}-\varepsilon_{\perp}) (\mathbf{n}\cdot\mathbf{E}^*)^2 -\varepsilon_0^* \varepsilon_{\perp} \mathbf{E}^*\cdot\mathbf{E}^*,\label{dielectric_density}\\
	&G^*_i = -\gamma_1^* N^*_i - \gamma_2^* e^*_{ik} n_k,\hspace{0.2cm}
	e^*_{ij} = \frac{1}{2}\left(\frac{\partial v^*_{i}}{\partial x^*_{j}}+ \frac{\partial v^*_{j}}{\partial x^*_i}\right),\label{body_force}\\
	&N^*_i = \dot{n}_i - \omega^*_{ik}n_k,\hspace{0.2cm} \omega^*_{ij} = \frac{1}{2}\left(\frac{\partial v^*_{i}}{\partial x^*_{j}}-\frac{\partial v^*_{j}}{\partial x^*_{i}} \right).
\end{eqnarray} 
Here, subscripts $i, j, k$ denote vector indices and the Einstein summation convention is assumed. The bulk free energy density $W^*$ consists of elastic and dielectric contributions. Nematic molecules prefer to align with their neighbors locally, a preference modeled by a bulk elastic (Frank) energy, $W_{\rm e}^*$, where in Eq.~(\ref{elastic_density}) the widely-used one-constant approximation is used \cite{DeGennes1995}. In addition, NLC molecules respond to an applied electric field $\mathbf{E}^*$, which induces a force causing them to align parallel or perpendicular to the field direction, giving rise to a dielectric free energy contribution modeled by $W_{\rm d}^*$ in Eq.~(\ref{dielectric_density}). The constant $\varepsilon_0^*$ is the permittivity of free space and $\varepsilon_{\parallel}$ and $\varepsilon_{\perp}$ are the relative dielectric permittivities parallel and perpendicular to the long axis of the nematic molecules. The molecular orientation induced by the field depends on the sign of the dielectric anisotropy, $\varepsilon_{\parallel}-\varepsilon_{\perp}$; in line with the most common situation we mostly assume parallel orientation, associated with $\varepsilon_{\parallel}-\varepsilon_{\perp}>0$. For our setup the electric field $E^* = \nabla^* \Psi^*$ is generated by applying a potential difference $V_0^*$ across the electrode pair; for modeling simplicity we follow the approach of Brown {\it et al.}~\cite{Brown2009} and approximate the piecewise continuous substrate potential by $\Psi^*(x^*,0,t^*)= V_0^*\cos(\frac{\pi x^*}{2d^*})$, where $d^*$ is the electrode spacing.
The parameters $\gamma_1^*$ and $\gamma_2^*$ in $\mathbf{G}^*$ are constant viscosities, while $e_{ij}^*$ and $\omega_{ij}^*$ are symmetric and antisymmetric rate-of-strain tensors for the material.  

The stress tensor $\mathbf{\sigma}^*$ for the NLC can be written as 
\begin{align}
 \mathbf{\sigma}^*= -p^* I +\sigma^{\rm e*}+\sigma^{\rm d*}+\sigma^{\rm v*},
\end{align}
where $\sigma^{\rm e*},\sigma^{\rm d*}, \sigma^{\rm v*}$ are the elastic, dielectric and viscous contributions, respectively, and are defined as
 \begin{eqnarray}
 &\sigma^{\rm e*}=-K^* \nabla^* \mathbf{n}\cdot (\nabla^* \mathbf{n})^T,  \label{def-sigma-e} \\
 &\sigma^{\rm d*}=\varepsilon_0^* (\nabla^*\Psi^* \otimes \nabla^*\Psi^*) \varepsilon (\mathbf{n}), \label{def-sigma-d}\\
 &(\sigma^{\rm v*})_{ij} = \alpha_1^* n_k n_p e^*_{kp} n_i n_j + \alpha_2^* N^*_{i}n_j + \alpha_3^* N^*_j n_i \nonumber \\
 &+\alpha_4^* e^*_{ij} + \alpha_5^* e^*_{ik} n_k n_j + \alpha_6^* e^*_{jk}, n_k n_i,
 \end{eqnarray}
where $\varepsilon (\mathbf{n}) = \varepsilon_{\perp} I +(\varepsilon_{\parallel}-\varepsilon_{\perp}) \mathbf{n}\otimes \mathbf{n}$ is the dielectric tensor, and the $\alpha_i^*$ are constant viscosities related to $\gamma_i^*$ in Eq.~(\ref{body_force}) by $\gamma_1^* = \alpha_3^*-\alpha_2^*, \gamma_2^* = \alpha_6^*-\alpha_5^*$ and to each other by the Onsager relation, $\alpha_2^* +\alpha_3^* = \alpha_6^* -\alpha_5^*$. 

 In addition to the electric field, NLC molecular orientation is sensitive to interactions with the bounding surface, a phenomenon known as anchoring.  At a solid substrate, anchoring is determined by the chemical interactions between the NLC and the substrate. It is a common practice in applications to treat the substrate (chemically or mechanically) to impose strong planar anchoring with respect to the surface, therefore we assume a Dirichlet condition $\theta(x^*,0,t^*) = \pi/2$ at the lower substrate. At a free surface, the director typically prefers to align normal to the surface (homeotropic anchoring), hence at $z^*=h^*(x^*,t^*)$ we impose weak homeotropic anchoring with associated anchoring strength $\scA^*$, modeled by a Rapini-Papoular surface energy contribution, \[\gamma^* -(\scA^*/2) (\mathbf{n}\cdot \mathbf{\nu})^2\] (where $\mathbf{\nu}$ is the unit outward normal to the free surface), to the total energy \cite{Rapini}. In experiments,
typical values for weak anchoring vary between $10^{-5}-10^{-6} \rm{Jm^{-2}}$ \cite{Mema2017}.
For the velocity $\mathbf{v^*}$, we assume no-slip and no-penetration conditions at the lower boundary, and a kinematic boundary condition together with a balance of normal and tangential stresses at the free surface. 

\subsection{Thin Film Model Derivation}

We employ standard long wave theory scalings to non-dimensionalize the governing equations:
\begin{eqnarray}
&x^* = d^*  x, \hspace{0.25cm} z^* = \delta d^* z, \hspace{0.25cm} u^* = U^* u,\hspace{0.25cm}w^* = \delta U^* w,
\nonumber \\ 
&t^* = \frac{d^*}{U^*} t,\hspace{0.25cm} p^* = \frac{\mu^* U^*}{\delta^2 d^*}p,\hspace{0.25cm} W^* = \frac{W K^*}{\delta^2 d^{*2}}.
\end{eqnarray}
Here, $d^*$ is the length scale of the electrode spacing along the $x^*$-axis, $U^*$ is the typical flow speed in the $x^*$-direction and $\mu^* = \alpha^*_4/2$ is the representative viscosity scaling in the pressure.

Experiments by Brown {\it et al.} \cite {Brown2009, Brown2010, Mottram2011} consist of a dielectric fluid film in which the typical film height $h_0^*$ is much smaller than the electrode spacing ($h_0^* \sim 15 {\rm \mu m}$ and $d^* \sim 120 {\rm \mu m}$), hence we set $\delta = h_0^*/d^*\ll 1$ to be the aspect ratio of the film.  Typical values of $\delta$ range from $0.05-0.5$ in the experiments~\cite{Brown2009, Brown2010}.  If the free surface in the dimensional variables is given by $z^* = h^*(x^*,t^*)$ then we write $h^* = h_0^* h$ and the dimensionless free surface representation is $z = h(x,t)$.

\subsubsection{Energetics}

Following an approach similar to that of Lin {\it et al.} \cite{TS-Lin} (which does not include electric field effects), provided the inverse Ericksen number $\scN = K^*/(\mu^* U^* d^*)$ and the dielectric parameter ${\cal D}$ (defined in Eq.~(\ref{W_nondim}) below) are both $O(1)$, the director energetics reduce to the Euler-Lagrange equations for minimizing total dimensionless free energy per unit length in the $x$-direction, given by
\begin{align}
J = \int_{0}^{h}\scN W(\theta,\theta_z, \Psi_z) \ dz + g(\theta)|_{z=h}, 
\end{align}
where $W$ is the total dimensionless bulk energy density and $g (\theta)$ is the Rapini-Papoular surface energy density, given in dimensionless form by
\begin{gather}
W = \frac{1}{2} \theta_z^2  -\scD \Psi_z^2 (\varpi + \cos^2 \theta),\label{W_nondim}\\
\scD = \frac{V_0^{*2} \varepsilon_0^* (\varepsilon_{\parallel}-\varepsilon_{\perp})}{2K^*}, \qquad \varpi = \frac{\varepsilon_{\perp}}{\varepsilon_{\parallel}-\varepsilon_{\perp}},\nonumber\\
g(\theta)|_{z=h} = \gamma - \frac{\scA}{2} ({\bf{n}} \cdot \nu)^2|_{z=h} =\gamma - \frac{\scA}{2} \cos^2 \theta |_{z=h}.
\label{RP-dimless}
\end{gather}
Here, $\scD$ represents the relative strength of the dielectric anisotropy and elasticity, $\gamma =\delta^2 d^{*2}\gamma^*/(K^*h_0^*)$ is the dimensionless surface tension coefficient defined in terms of the dimensional surface tension $\gamma^*$ and $\scA = \delta \scA^*d^*/K^*$ is the dimensionless anchoring strength at the free surface.

We follow the approach described by Cummings {\it et al.}~\cite{nonuniform_field} to minimize the total free energy density of the layer with respect to variations in $\theta$ and $\Psi$. The first variations must vanish at an extremum and the signs of the second variations indicate whether an energy minimum is reached.  The bulk terms lead to the Euler-Lagrange equations,
\begin{align}
\theta_{zz} -\scD \Psi_z^2\sin 2\theta = 0, \label{energy-x}\\
\scD \bigl( \Psi_z (\varpi+\cos^2\theta) \bigr)_z=0,
\label{energy-z}
\end{align} 
while the surface terms lead to one term that can be eliminated, giving the weak anchoring condition,
\begin{align*}
W_{\theta_z}&=g_{\theta}\quad\mbox{on \quad $z=h$}.
\end{align*}
The remaining boundary conditions on $\theta$ and $\Psi$ are Dirichlet conditions: strong planar anchoring on $z=0$, specified surface potential $\Psi (x,0,t)$, and we assume the free upper surface is electrically insulated (for simplicity), giving the following complete set of boundary conditions for $\theta$ and $\Psi$:
\begin{align}
\theta =\pi/2\qquad &{\rm on} \quad z =0,
\label{SA}\\
\theta_z + \frac{\scA}{2}\sin 2\theta = 0 \qquad &{\rm on} \quad z = h,
\label{WA}\\
\Psi =  \cos\left(\frac{\pi}{2} x \right) \qquad &{\rm on} \quad z = 0, \label{SP}\\
\Psi= 0\qquad &{\rm on} \quad z = h.
\label{IBC}
\end{align}
Although both $\theta$ and $\Psi$ are functions of the three independent variables $x,z,t$, in much of the following we will suppress the explicit $t$-dependence due to the quasistatic nature of the boundary value problem these functions satisfy. We will also, where convenient, suppress the independent variables altogether in $\theta$ and $\Psi$ (likewise in $h$). 

\subsubsection{Momentum Equations}

We use the pressure scale $\mu^* U^*/(\delta^2 d^*)$ to non-dimensionalize the stress tensor $\sigma^*$. Then, provided that the inverse Ericksen number $K^*/(\mu^* U^* d^*) = O(1)$, the momentum equations reduce to the condition that the stress tensor be divergence-free, and the equations governing the fluid flow can be extracted by retaining the leading order terms in
\begin{align}
\delta \frac{\partial \sigma_{11}}{\partial x} + \frac{\partial \sigma_{13}}{\partial z}=0,\\
\delta \frac{\partial \sigma_{31}}{\partial x} + \frac{\partial \sigma_{33}}{\partial z}=0,
\end{align}
where subscripts 1 and 3 refer to the $x$ and $z$ coordinate directions respectively, in our 2D geometry. As a result, the leading-order equations are found to be 
\begin{align}
p_x &=2\scD\scN \bigl(\Psi_x\Psi_z (\varpi +\cos^2\theta )\bigr)_z \label{x-mom}\\
&+ \bigl( A_1(\theta) u_z \bigr)_z-\scN \bigl(\theta_x\theta_z \bigr)_z,\nonumber\\
p_z &= 0,\label{z-mom}
\end{align}
for the $x$ and $z$ components respectively, where 
\begin{align}
A_1 (\theta) &= 1+ (\alpha_5 - \alpha_2) \cos^2 \theta + 2 \alpha_1 \sin^2\theta \cos^2 \theta \nonumber\\
&+ (\alpha_3 + \alpha_6) \sin^2\theta. 
\end{align}

Following the approach taken by Lin {\it et al.} \cite{TS-Lin,TS-Lin2013Note}, 
 the normal and tangential stress balance conditions may be derived. The normal stress is balanced by the surface tension contribution in the Young-Laplace condition, which yields
 \begin{equation}
p+2\scN W =-\scC h_{xx} \quad \mbox{on $z=h(x,t)$,}
\label{normal-stress}
\end{equation}
  where $\scC= \delta^3 \gamma^*/(\mu^* U^*)$ is an inverse capillary number. The tangential stress is balanced by the surface energy gradients along the free surface, which (after some algebra) reduces to the condition
  \begin{align}
  u_z=0 \quad \mbox{on $z=h(x,t)$.}
  \label{SBC}
  \end{align}
Equation~(\ref{z-mom}) together with the normal stress condition of Eq.~(\ref{normal-stress}) gives the pressure $p(x,t)$ throughout the layer as
\begin{equation}
p=-2\scN W|_{z=h}-\scC h_{xx},
\label{p-sol}
\end{equation}
 while Eq.~(\ref{x-mom}) together with the tangential stress condition in Eq.~(\ref{SBC}) gives the velocity gradient across the layer as 
\begin{eqnarray}
&A_1(\theta) u_z =\scN \left[ \theta_x(x,z)\theta_z(x,z)-\frac{\scA}{2} \theta_x(x,h)\sin 2\theta(x,h) \right]\nonumber\\
			   &-2\scD\scN \Psi_z(x,z)\left[\varpi +\cos^2\theta(x,z)\right] \left[\Psi_x(x,z)-\Psi_x (x,h)\right]\nonumber\\
				&-2\scN (z-h) \left[W_x+h_x W_z \right]|_{z=h} -\scC (z-h) h_{xxx}.\hspace{1.2cm}\label{u_z}
\end{eqnarray}
Imposing the no-slip and no-penetration boundary conditions, $u = v = 0$ at $z = 0$, and the kinematic boundary condition at $z = h(x,t)$, together with mass conservation and the integral relation $	\int_0^h u \ dz = \int_0^h u_z(h-z) \ dz$, gives
\begin{eqnarray}
\frac{\partial h}{\partial t} + \frac{\partial}{\partial  x} \left(\int_0^{h} u_z (h-z) \  dz \right)= 0. \label{intermediate}
\end{eqnarray} 
Substituting Eq.~(\ref{u_z}) into Eq.~(\ref{intermediate}) above, we obtain a partial differential equation (PDE) that governs the evolution of the NLC film height $h$, 
\begin{eqnarray}
&h_t + \scC \frac{\partial}{\partial x}\left(h_{xxx} \int_0^h \frac{(h-z)^2}{A_1(\theta)} dz \right)\label{gov_eqn} \\
&+\scN \frac{\partial}{\partial x}\left(\int_0^h 
\frac{(h-z)[T_1(x,z)+T_2(x,h)+T_3(x,h)(h-z)]}{A_1(\theta)} dz\right) = 0,\nonumber
\end{eqnarray}  
where the functions $T_1(x,z)$, $T_2(x,h)$ and $T_3(x,h)$ take the following forms: 
\begin{align}
T_1(x,z) &= \theta_x(x,z)\theta_z(x,z)\label{T1}\\
&-2\scD \Psi_x(x,z) \Psi_z(x,z)[\varpi + \cos^2(\theta(x,z))]\nonumber,
\end{align}
\begin{align}
T_2(x,h) &= -\frac{\scA}{2}\theta_x(x,h)\sin(2\theta(x,h))\label{T2}\\
&+ 2\scD \Psi_x(x,h) \Psi_z(x,h)
[\varpi + \cos^2(\theta(x,h))]\nonumber,
\end{align}
\begin{align}
T_3(x,h) &= \scA
\sin(2\theta(x,h))[\theta_{zx}(x,h) + h_x \theta_{zz}(x,h)]\nonumber\\
 &+ 2\scD \Psi_{z}^2(x,h)\sin 2\theta(x,h)[\theta_x(x,h)+h_x \theta_z(x,h)]\nonumber\\
 &- 4\scD \Psi_z(x,h)(\varpi+\cos^2(\theta(x,h)))\nonumber\\
 &\times [\Psi_{zx}(x,h)+h_x \Psi_{zz}(x,h)].\label{T3}
\end{align}
In order to simulate a thin film of infinite lateral extent overlying a periodic array of electrodes, 
we impose periodic boundary conditions in the $x$-direction as follows: 
\begin{align}
&h(0,t) = h(L,t), \quad h_x(0,t) = h_x(L,t), \label{BC_x}\\
&h_{xx}(0,t) = h_{xx}(L,t), \quad h_{xxx}(0,t) = h_{xxx}(L,t),\nonumber
\end{align}
where the length of the computational domain $L$ is taken to be an even integer (corresponding to simulating flow over an integer number of electrode pairs). We use $L=4$ for all simulations presented in this paper, representing a single period-unit of the interdigitated electrode setup.

\subsection{Solution Scheme \& Discretization}

Equation~(\ref{gov_eqn}), coupled with the boundary-value system Eqs.~(\ref{energy-x})--(\ref{energy-z}), subject to the boundary conditions given by Eqs.~(\ref{BC_x}) and Eqs.~(\ref{SA})--(\ref{IBC}) respectively, describe the flow of a NLC film driven by the nonuniform electric field generated by the interdigitated substrate electrodes. 

Solving these equations simultaneously, even for the 2D geometry considered here, poses a significant computational challenge. Since the boundary-value problem (BVP) Eqs.~(\ref{energy-x})--(\ref{IBC}) is quasistatic (reflecting the assumptions on the much slower time scale of the flow relative to the timescales on which NLC molecules and electric field respond to changes in film geometry) we are able to solve it independently for a dense grid of film heights $h$.  
To do this, we rewrite the BVP as a vector system of four first-order ordinary differential equations for $\theta$, $\theta_z$, $\Psi$, $\Psi_z$ and apply the MATLAB routine {\tt bvp4c}.  An initial guess [$\theta_0,\Psi_0$] is required to start the routine; for this we use the results of the analogous uniform field problem, obtained as outlined by Mema {\it et al.} \cite{nonuniform_field,Mema2015}, for each height $h$. The procedure builds a ``library'' that consists of the director configuration $\theta(x,z)$ and electric potential $\Psi(x,z)$ for a range of film heights $0.01\leq h \leq 2.00$ with $\Delta h = 0.01$, as an electric potential $\Psi(x) = \cos(\frac{\pi}{2}x)$ is applied at $z=0$ for $0\leq x \leq 4$ with $\Delta x = 0.02$ (the value of $\Delta z$ to compute the solutions $\theta,\Psi$ for each pair $(x,h)$ is selected by the routine). 

Once the BVP is solved for all discrete film heights $h\in [0.01, 2.0]$ and all discrete $x\in [0,4]$, we solve Eq.~(\ref{gov_eqn}) numerically. The integrals in Eq.~(\ref{gov_eqn}) are difficult to evaluate directly; in the interests of retaining maximum tractability we follow Lin {\it et al.} \cite{TS-Lin,Lam2015incline} in using a two-point trapezoidal rule to approximate them and hence obtain the following fourth-order nonlinear partial differential equation for the film thickness $h(x,t)$,
\begin{align}
\frac{\partial h}{\partial t}
&+ \mscC\frac{\partial}{\partial x}[h^3 h_{xxx}]\label{new_gov_eqn}\\
&+\mscN \frac{\partial}{\partial x}[h^2 (T_1(x,0)+T_2(x,h))+h^3 T_3(x,h)]=0,\nonumber
\end{align}
 where $\mscC = \scC/(2(1+\alpha_3+\alpha_6))$, $\mscN = \scN/(2(1+\alpha_3+\alpha_6))$. The arguments of $h$ are suppressed in Eq.~(\ref{new_gov_eqn}) and those for $T_i$ are understood to be $(x,z)$; the quantities   $T_1(x,0),T_2(x,h),T_3(x,h)$ are then given by Eqs.~(\ref{T1})-(\ref{T3}) respectively.  
We use a finite difference method based on central differences to discretize the spatial terms in Eq.~(\ref{new_gov_eqn}) and use the built-in MATLAB routine {\tt ode15s}, which is a variable-step variable-order solver \cite{SIAM_ODE15s}, to approximate the height of the film at each time step. The procedure used to obtain the numerical results is as follows: given an initial film profile, $h_0(x)$ we solve Eq.~(\ref{new_gov_eqn}) at each time step, extracting the director and electric potential configurations in the process as necessary from our previously built library. 

\section{Results}

In this section we use the above-derived model comprising Eq.~(\ref{new_gov_eqn}) coupled with Eqs.~(\ref{energy-x})--(\ref{IBC}) to investigate the flow of an initially flat film in the presence of a nonuniform electric field. We focus on the evolution of the free surface height $h(x,t)$, as well as the director and electric fields within the layer, as we vary the initial film height, $h_0$ and the anchoring conditions at the upper bounding surface. With a view to later qualitative comparison with the experiments of Brown {\it et al.}~\cite{Brown2010,Brown2009,Brown2011} we explore a range of initial film heights: $h_0 = 1.5$ is taken to be representative of a thick film in our model, $h_0 = 1.0$ an intermediate film and $h_0\leq 0.5$ is considered a thin film. Depending on the thickness of the NLC film, dimensionless weak anchoring strength can lie in the approximate range $\scA \in (0.125,50)$. Consistent with this range of values, in our simulations
we use values $\scA = 1,10,50$ to represent a range of weak anchoring conditions.

Figures~\ref{fig_h:h=1.0_A=10}--\ref{h:h=1.5_A=10} present results for evolution of a NLC film with strong planar anchoring at $z = 0$ and weak homeotropic anchoring of strength $\scA = 10$ at the free surface $z = h(x,t)$, for several different values of $h_0$. Figures~\ref{fig:A=1_h0=1.0}--\ref{fig:A=50_h0=1.5} illustrate the evolution for lower ($\scA = 1$) and higher ($\scA = 50$) values of the free surface anchoring strength. Figures~\ref{fig:h=1.0_A=10_planar}-\ref{fig:h=0.3_A=10_planar} show film evolution for strong planar substrate anchoring and weak planar free surface anchoring of strength $\scA = 10$ and negative dielectric anisotropy $\scD <0$.  In all cases, the layer is subjected to a dimensionless electric potential $\Psi(x,0) = \cos(\frac{\pi x}{2})$ on the substrate, and an insulating boundary condition $\Psi(x,h) = 0$ at the free surface. We set the inverse capillary number and Ericksen number to $\mathscr{C} = \mathscr{N}= 0.625$ and the relative strength of the dielectric anisotropy $\scD = \pm 10$ throughout. Our simulations are stopped either when the film height $h$ falls below $0.1$ or when a steady state is reached. 

 In many of our simulation plots, we superimpose representative snapshots of the director field as short directed line segments. We note that our model is invariant under the transformation $\theta\to -\theta$ (a limitation of our restriction to a 2D thin film model), hence a choice must be made when selecting $\theta$. Although numerically our code selects values continuously in the range $\theta\in [0,\pi/2]$, our plots show line segment representations of the director that align visually with the electric potential lines (there is no mathematical distinction between the two possible states in our model).

\subsection{Effect of initial film height on free surface evolution}

\begin{figure}[h!]
 	\subfigure[]{
		\includegraphics[width=0.5\textwidth]{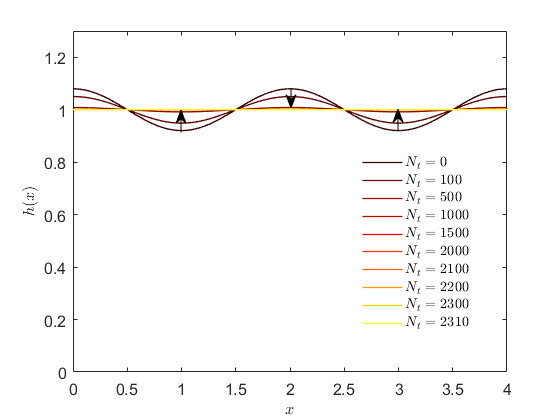}}
	\subfigure[]{\includegraphics[width=0.5\textwidth]{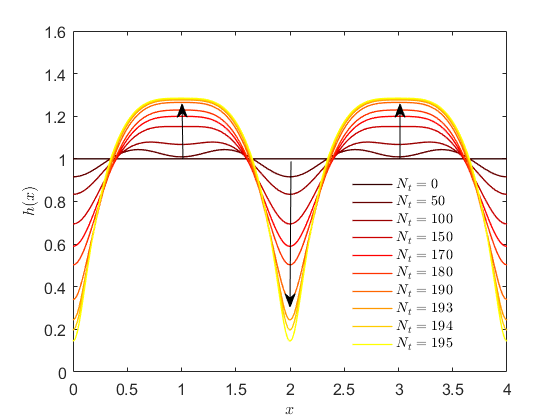}\label{fig:D=0}}
		\caption{(a)~Evolution of the surface height for a perturbed film of initial height $h(x) =1.0 + \frac{1}{4 \pi}\cos(\pi x)$, anchoring strength $\scA= 10$, $\mathscr{C}=\mathscr{N}= 0.625$ when no electric field is applied: $\scD = 0$ for various times $t = N_t/N$ where $N = 8000$ and $N_t$ is specified in the legend. (b) Evolution of the free surface height of a flat film with initial height $h_0 = 1.0$, $\scA= 10$, $\mathscr{C} = \mathscr{N}=0.625$ and $\scD = 10$ for various times $t = N_t/N$ where $N = 8000$ and $N_t$ is specified in the legend. }
		\label{fig_h:h=1.0_A=10}
	\end{figure}

We consider NLC films subject to strong planar substrate anchoring $\theta(x,0) = \pi/2$, and weak homeotropic anchoring at $z=h(x,t)$ with dimensionless anchoring strength $\scA=10$. We begin by demonstrating film stability in the absence of an applied electric field. As can be seen in Fig.~\ref{fig_h:h=1.0_A=10}(a), a slightly perturbed film of initial thickness $h_0(x) = 1.0+ \frac{1}{4\pi} \cos(\pi x)$ flattens in this case, demonstrating the stability of a flat film to perturbations. With no external field the NLC molecules align with their neighbors within the layer, while respecting the anchoring conditions at the boundaries. 

\begin{figure}[h!]
	{\center
\hspace*{-0.6in}
	\includegraphics[width=1.2\linewidth]{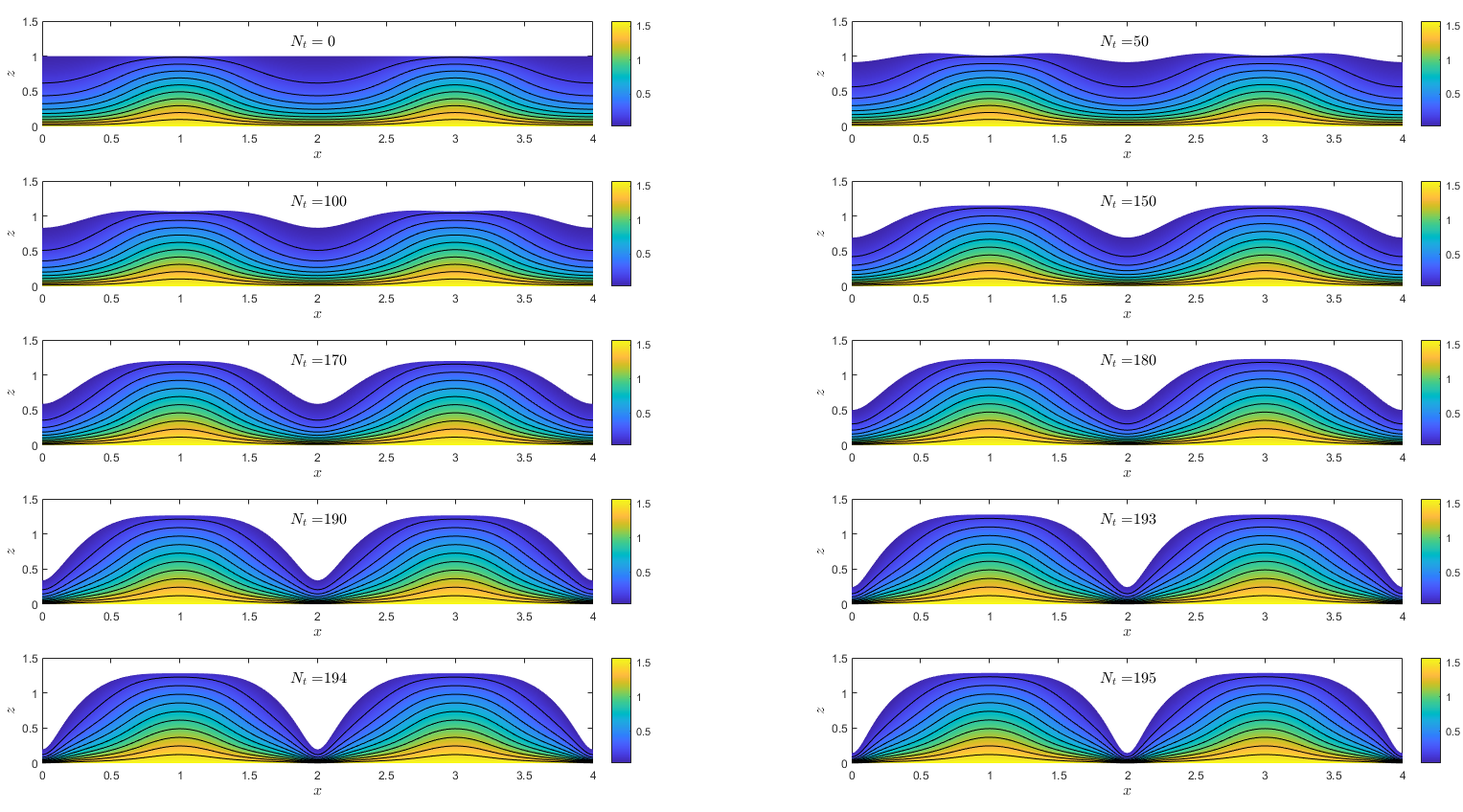}
 		\caption{Evolution of the director field $\theta(x,z)$ and its level curves for the film of Fig.~\ref{fig_h:h=1.0_A=10}(a).
 		\label{theta:h=1.0_A=10}}}
	\end{figure}

 We next consider a flat film, of initial dimensionless height $h_0 = 1.0$ (intermediate thickness), on application of the nonuniform electric field. In this case the film rapidly deforms, as seen in Fig.~\ref{fig_h:h=1.0_A=10}(b). The fluid collects around $x = 1$ and $x=3$ (points of zero substrate potential) and begins to thin around $x = 0,2$ and $4$ (where the substrate potential takes values $+1,-1,+1$, respectively). In the presence of the nonuniform potential the polar nature of the NLC molecules induces a force that tends to align them parallel to the local electric field, competing with the forces due to the internal elasticity and the anchoring boundary conditions at $z = 0,h$ respectively. As a result, molecules begin to migrate from the regions directly above the electrodes, $x=0,2$ and $x = 4$ where $\Psi(x,0) = \pm 1$ and $\Psi_z(x,0)$ is large, towards the inter-electrode regions above $x= 1$ and $x = 3$ where the electric potential $\Psi(x,0)$ and its gradient $\Psi_z(x,0)$ are both small. 
	\begin{figure}[h!]
{\center
\hspace*{-0.6in}
	\includegraphics[width=1.2\linewidth]{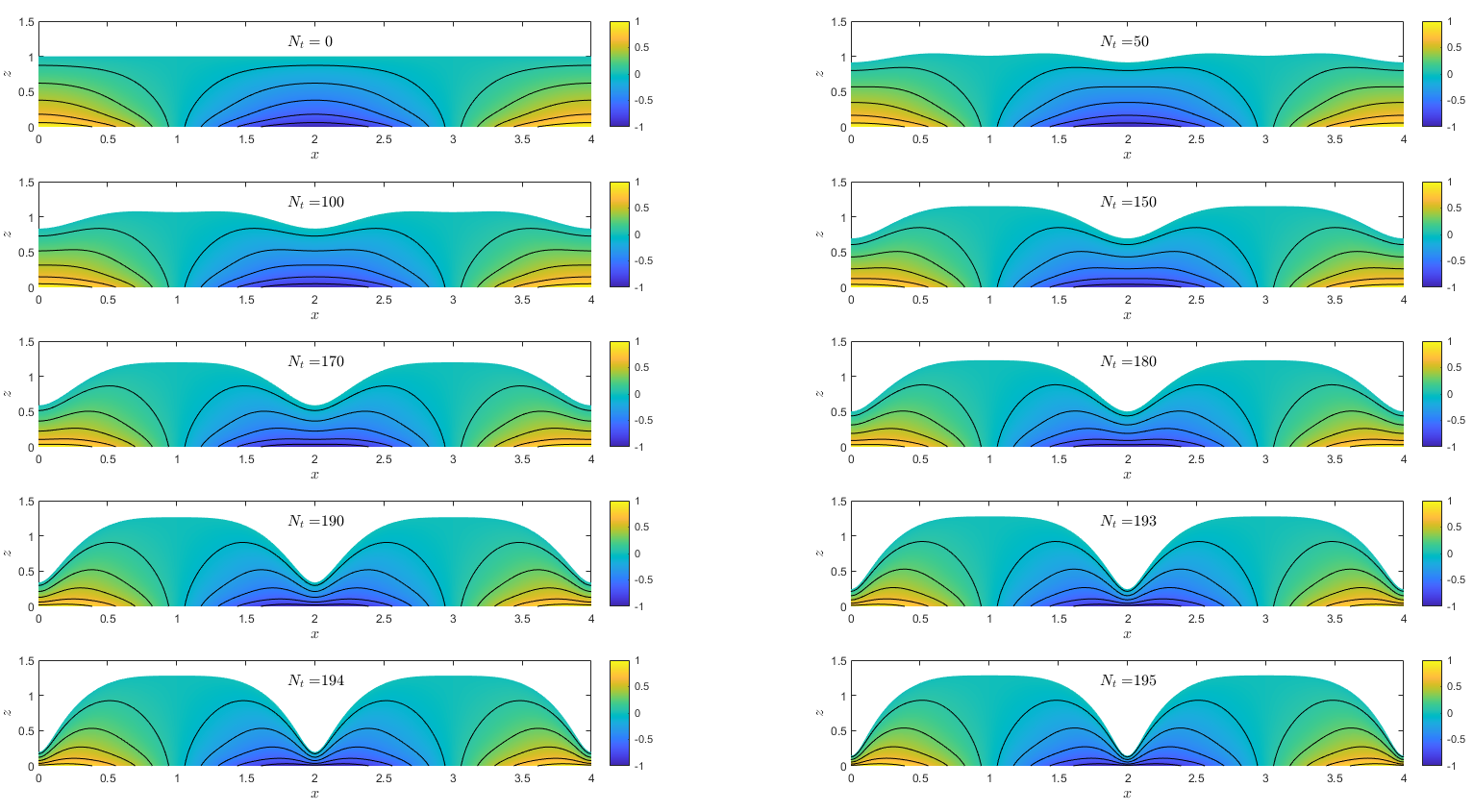}
	\caption{Evolution of the electric potential $\Psi(x,z)$ and its level curves for the film of Fig.~\ref{fig_h:h=1.0_A=10}(a).
	\label{pzi:h=1.0_A=10}}}
\end{figure}
This may be seen in Fig.~\ref{pzi:h=1.0_A=10}, which illustrates the electric potential $\Psi(x,z)$ and its level curves within the NLC layer at the same times shown in Fig.~\ref{fig_h:h=1.0_A=10}.
Figure~\ref{theta:h=1.0_A=10} shows the corresponding behavior of the director field: the director angle $\theta(x,z)$ and its level curves are plotted within the layer. We see that the director gradient $|\theta_z|$ is large in the regions above the electrodes, $x = 0,2$ and $x = 4$, and smaller in the regions between electrodes, at $x = 1$ and $x = 3$.

This behavior is qualitatively consistent with the experimental results reported by Brown {\it et al.}~\cite{Brown2010,Brown2009,Mottram2011} where a similar set up is considered: a layer of an isotropic dielectric fluid (1-decanol oil), rather than NLC, coats a glass substrate on which interdigitated ITO (indium tin oxide) electrodes are patterned. The authors observe that liquid flows from regions where the electric potential gradients are small to those regions where gradients are highest. We will briefly address ways in which the specific nematic nature of the film in our study may impact results later, when we consider how the details of the anchoring conditions affect the surface height evolution.  
   
\begin{figure}[h!]
	{\center
\hspace*{-0.2in}
\subfigure[]{
	\includegraphics[width=0.5\textwidth]{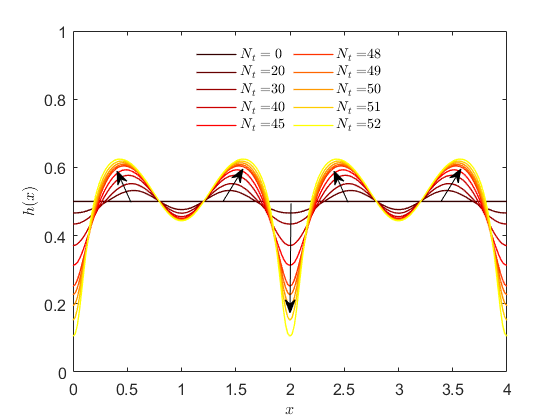}}}
\subfigure[]{
    \includegraphics[width=0.5\textwidth]{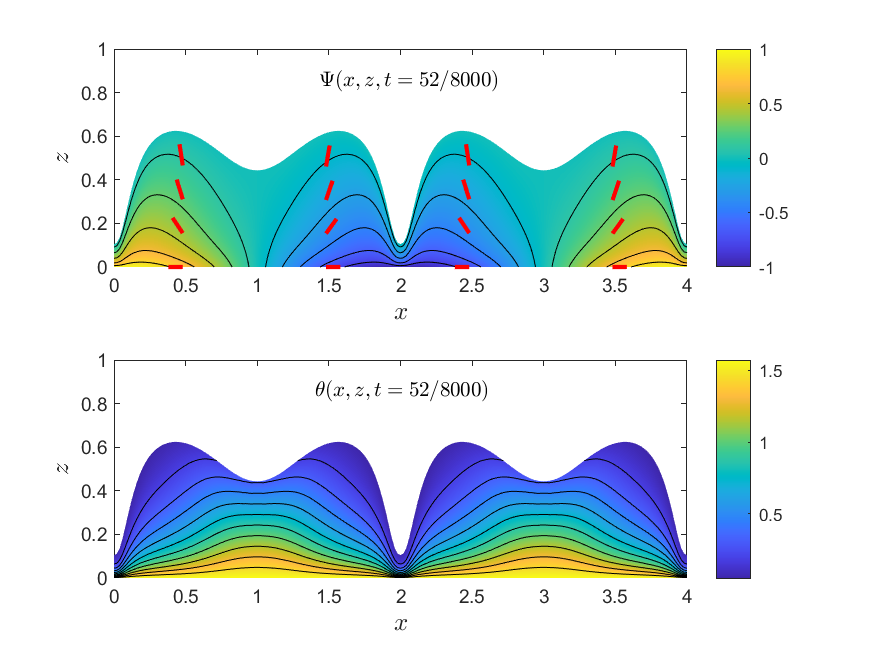}}
	\caption{(a)~Evolution of the free surface height of a film with initial height $h_0 = 0.5$ for $\scA= 10$, $\mathscr{C} =  \mathscr{N}=0.625$ and $\scD = 10$  for various times $t = N_t/N$ where $N = 8000$ and $N_t$ is specified in the legend. (b)~Electric potential $\Psi(x,z)$ and director field $\theta(x,z)$ at final time: $t = 52/N$ with director configuration shown in red.}
	\label{h:h=0.5_A=10}
\end{figure} 

We now investigate how changing the initial film height affects evolution. We first consider a thinner film of initial dimensionless thickness $h_0 = 0.5$, then a thicker film with $h_0 = 1.5$, both initially flat. Figure~\ref{h:h=0.5_A=10} presents the evolution of the thinner film in time.  For early times we observe the formation of wrinkling patterns, where the film thickens at $x = 0.5,1.5,2.5$ and $x = 3.5$ (representing the edges of the electrodes in our simple model); and thins at $x = 0,2,4$ (electrode midpoints) and at $x = 1,3$ (points mid-way between adjacent electrodes). As time progresses the film continues to thin significantly at the electrode midpoints $x=0,2,4$ where the potential gradient is large; these are the global film minima. Between the electrodes, where the potential gradient is small, further thinning of the film is suppressed and local minima persist at $x = 1$ and $x = 3$. We halt our simulations when the layer becomes too thin at the global minima and the director field no longer responds to the electric field in the interior of the layer. If the simulations are pushed further, solutions for the director configuration become unreliable. The evolution of the electric potential and the director field is qualitatively similar to the intermediate film height case: both director and electric potential gradients are large in the regions above the electrodes ($x = 0,2,4$) and small between them ($x = 1$ and $x = 3$). However, the gradients are larger for thinner films.  

\begin{figure}[h!]
	{\center
\hspace*{-0.2in}
\subfigure[]{
	\includegraphics[width=0.5\textwidth]{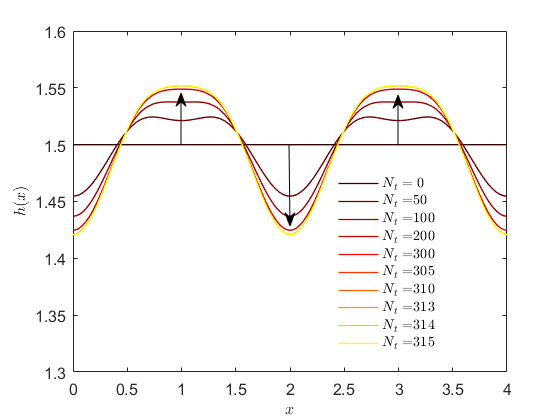}}}
\subfigure[]{
\includegraphics[width=0.5\textwidth]{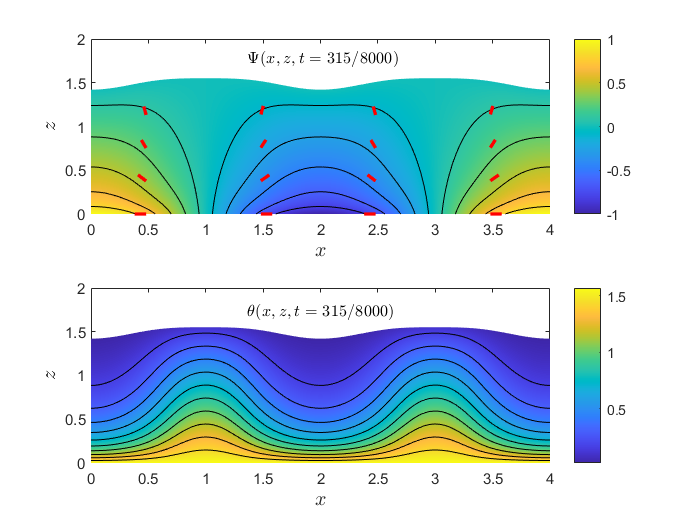}}
	\caption{(a)~Evolution of the free surface height of a film with initial height $h_0 = 1.5$ for $\scA= 10$, $\mathscr{C} = \mathscr{N}=0.625$ and $\scD = 10$ for various times $t = N_t/N$ where $N = 8000$ and $N_t$ is specified in the legend. (b)~Electric field potential $\Psi(x,z)$ and director field $\theta(x,z)$ at final time: $t = 315/N$ with director configuration shown in red.}
	\label{h:h=1.5_A=10}
\end{figure} 

 Figure~\ref{h:h=1.5_A=10} shows the free surface evolution of the thicker NLC film, $h_0 = 1.5$. For early times the evolution is similar to the intermediate film case $h
_0=1.0$ (Fig.~\ref{fig_h:h=1.0_A=10}), with the fluid collecting in regions where the electric potential gradients are small. As time progresses, however, the thicker film reaches a steady state, with approximately sinusoidal profile, exhibiting local minima of height $h \approx 1.42$ at electrode midpoints $x = 0,2$ and $4$ and local maxima of height $h \approx 1.55$ at $x = 1$ and $x = 3$, midway between electrodes. Our simulations stop when $|h(x,t_n)-h(x,t_{n-1})|<10^{-5}$, as we deem a steady state to have been reached.

\subsection{Effect of anchoring conditions on the evolution of surface height}

We next investigate the influence of different anchoring conditions on the evolution of an initially flat film. Specifically, we first study how, in the model described above, the strength of the weak homeotropic anchoring at the free surface affects the surface height evolution. Then, with a view to conducting simulations that are closer in spirit to the isotropic dielectric case studied by Brown and co-workers~\cite{Brown2009,Brown2010,Brown2011}, we briefly study the evolution of NLC films with negative dielectric anisotropy, subject to planar anchoring at both boundaries. 

\subsubsection{Weak homeotropic anchoring strength}

We consider the model described above, comprising Eq.~(\ref{new_gov_eqn}) coupled with Eqs.~(\ref{energy-x})--(\ref{IBC}), with strong planar anchoring at the lower substrate and weak homeotropic anchoring at the upper free surface. All previous simulations were for dimensionless (Rapini-Papoular) anchoring strength $\scA = 10$; we now simulate initially flat films where $\scA = 1$ and $\scA = 50$ at the free surface. We again compare results for three different initial film thicknesses, $h_0= 0.5,1.0$ and $1.5$.

\begin{figure}[h!]
 	{\center
\hspace*{-0.2in}
    \subfigure[]{
    \includegraphics[width=0.5\textwidth]{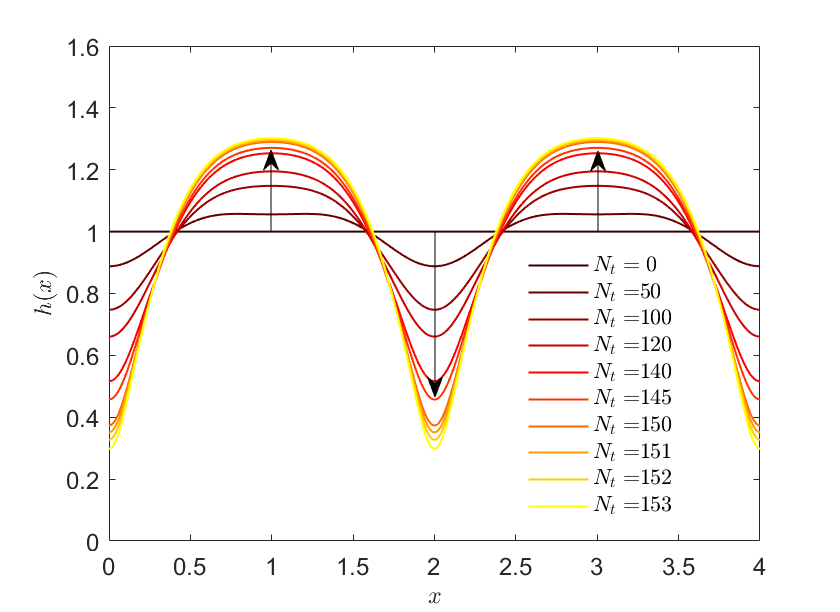}\label{fig:h=1.0_A=1}}}
    \subfigure[]{
    \includegraphics[width=0.5\textwidth]{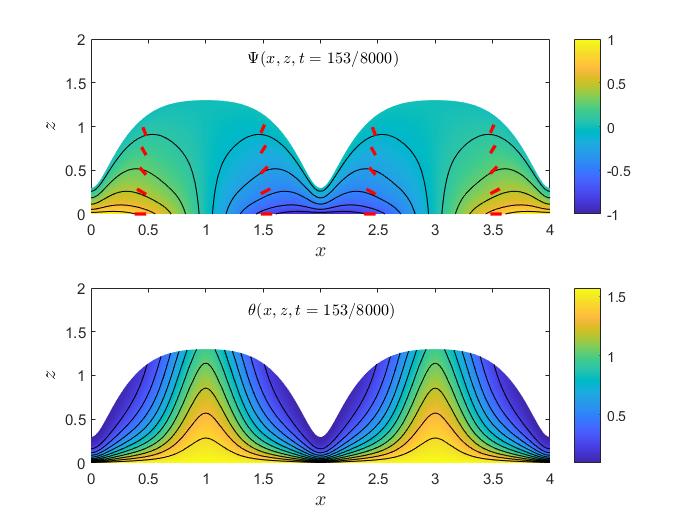}\label{fig:dir=1.0_A=1}}
    \caption{Evolution of an initially flat film, $h_0 =1.0$ with weak homeotropic free surface anchoring of dimensionless strength $\scA = 1$ and parameters: $\mathscr{C}=\mathscr{N}=0.625$ and $\scD = 10$.  Various times $t = N_t/N$ are shown in (a) where $N = 8000$ and $N_t$ is specified in the legend. (b)~Electric potential $\Psi(x,z)$ and director field $\theta(x,z)$ at final time $t =153/N$ with director configuration shown in red.}
    \label{fig:A=1_h0=1.0}
\end{figure}

\begin{figure}[h!]
	{\center
\hspace*{-0.2in}
    \subfigure[]{
    \includegraphics[width=0.5\textwidth]{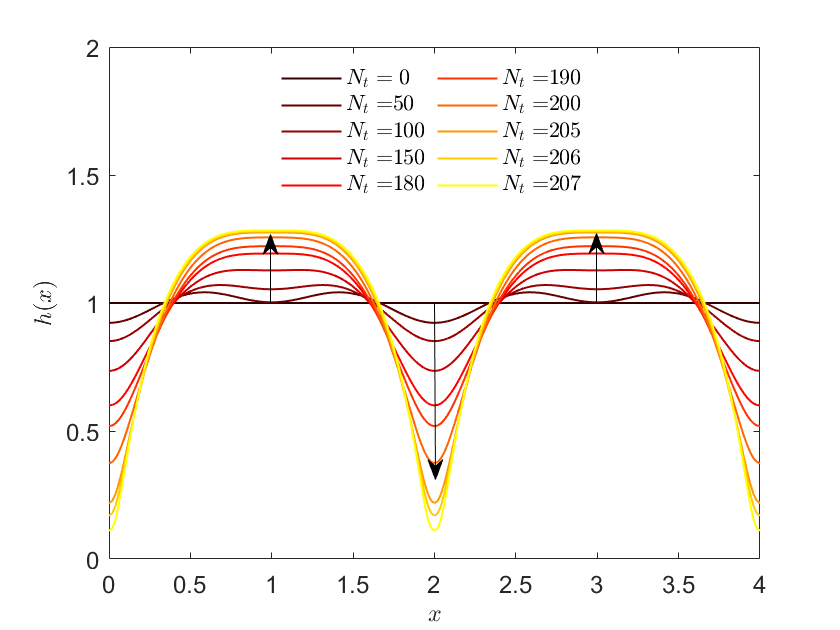}\label{fig:h=1.0_A=50}}}
    \subfigure[]{
    \includegraphics[width=0.5\textwidth]{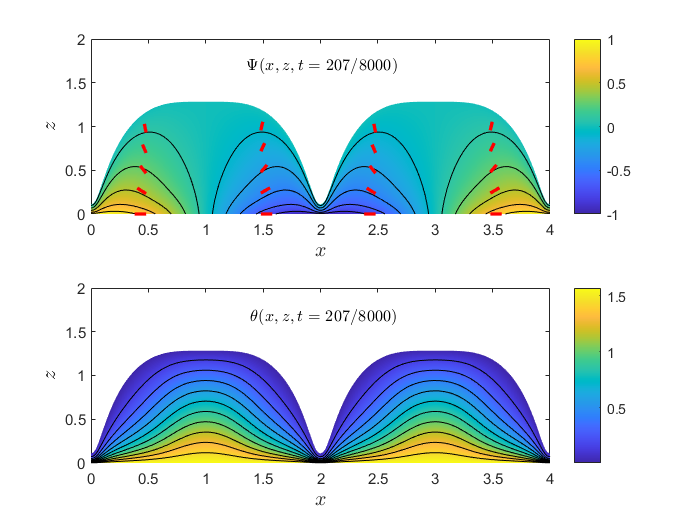}\label{fig:dir=1.0_A=50}}
    \caption{Evolution of an initially flat film, $h_0 =1.0$ with weak homeotropic free surface anchoring of dimensionless strength $\scA = 50$ and parameters: $\mathscr{C}=\mathscr{N}=0.625$ and $\scD = 10$.  Various times $t = N_t/N$ are shown in (a) where $N = 8000$ and $N_t$ is specified in the legend. (b)~Electric potential $\Psi(x,z)$ and director field $\theta(x,z)$ at final time $t =207/N$ with director configuration shown in red.}
    \label{fig:A=50_h0=1.0}
\end{figure}

 Figures~\ref{fig:A=1_h0=1.0} and \ref{fig:A=50_h0=1.0} show the evolution of an initially flat film of height $h_0 = 1.0$ and the corresponding electric potential $\Psi(x,z)$ and director field $\theta(x,z)$ at the final computed time for anchoring strengths $\scA = 1$ and $\scA = 50$ respectively. The evolution of the corresponding film with homeotropic anchoring of dimensionless strength $\scA = 10$ at the free surface was shown in Fig.~\ref{fig_h:h=1.0_A=10}(a). Similar to the $\scA = 10$ case, our simulations stop when the layer becomes too thin and the solution for the director field becomes unreliable. 
  While the evolution of the film is similar for all three anchoring strengths considered, there are some differences: first, in the weaker anchoring case (dimensionless anchoring strength $\scA = 1$), the director at the free surface deviates significantly from the preferred homeotropic orientation (see Fig.~\ref{fig:dir=1.0_A=1}); this deviation is especially apparent in regions between the electrodes (where the electric potential is close to zero). For (relatively) strong anchoring $\scA = 50$, by contrast, the director almost perfectly respects the preferred anchoring conditions at the free surface: see Fig.~\ref{fig:dir=1.0_A=50}.
  Additionally, we note that the maxima in the free surface height profile at $x = 1$ and $x = 3$ become more diffuse as the anchoring strength increases (peaks shown in   Fig.~\ref{fig:h=1.0_A=50} are wider than those in Fig.~\ref{fig:h=1.0_A=1}). 


   \begin{figure}[ht]
   	{\center
\hspace*{-0.2in}
    \subfigure[]{
    \includegraphics[width=0.5\textwidth]{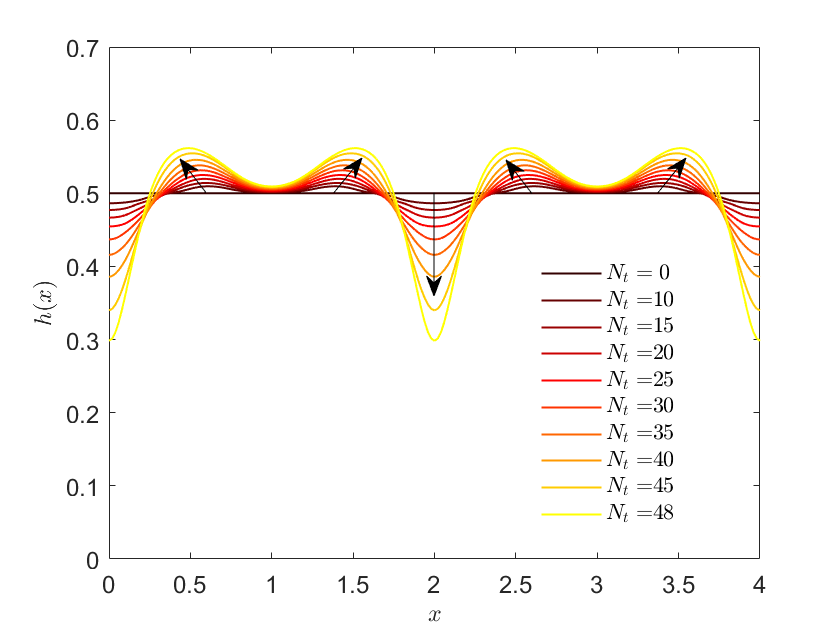}\label{fig:h=0p5_A=1}}}
    \subfigure[]{
    \includegraphics[width=0.5\textwidth]{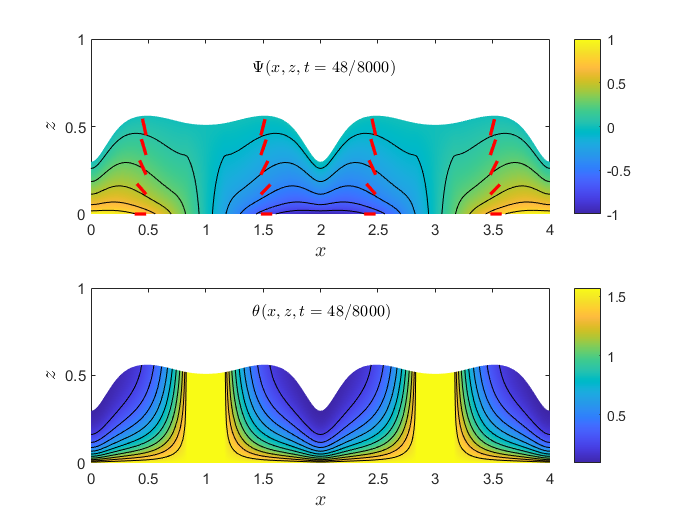}\label{fig:dir=0.5_A=1}}
    \caption{Evolution of an initially flat film, $h_0 =0.5$ with weak homeotropic free surface anchoring of dimensionless strength $\scA = 1$ and parameters: $\mathscr{C}=\mathscr{N}=0.625$ and $\scD = 10$.  Various times $t = N_t/N$ are shown in (a) where $N = 8000$ and $N_t$ is specified in the legend. (b)~Electric potential $\Psi(x,z)$ and director field $\theta(x,z)$ at final time $t =48/N$ with director configuration shown in red.}
    \label{fig:A=1_h0=0.5}
\end{figure}
\begin{figure}[ht]
	{\center
\hspace*{-0.2in}
    \subfigure[]{
    \includegraphics[width=0.5\textwidth]{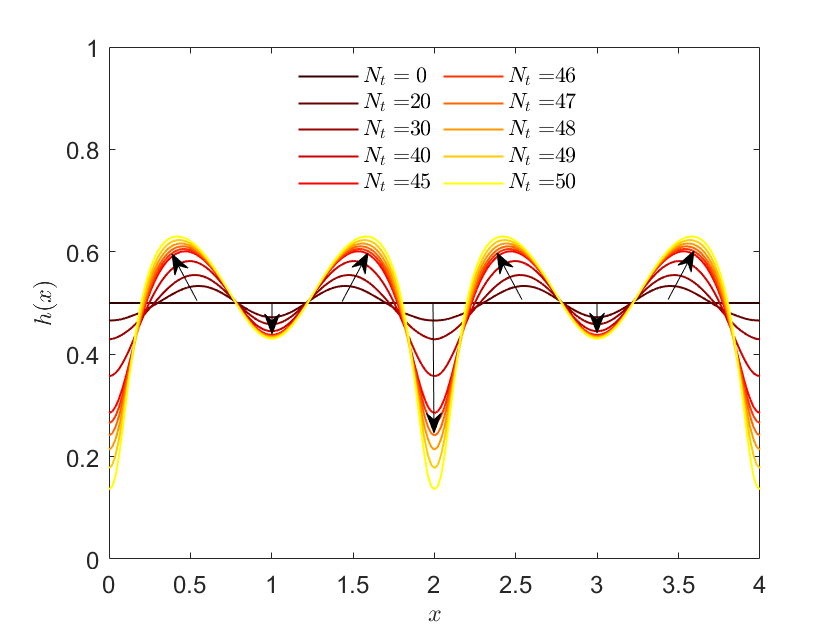}\label{fig:h=0.5_A=50}}}
    \subfigure[]{
    \includegraphics[width=0.5\textwidth]{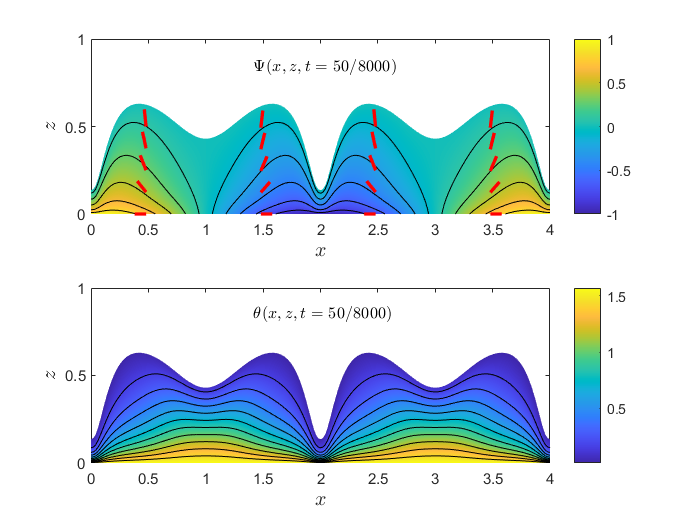}\label{fig:dir=0.5_A=50}}
    \caption{Evolution of an initially flat film, $h_0 =0.5$ with weak homeotropic free surface anchoring of dimensionless strength $\scA = 50$ and parameters: $\mathscr{C}=\mathscr{N}=0.625$ and $\scD = 10$.  Various times $t = N_t/N$ are shown in (a) where $N = 8000$ and $N_t$ is specified in the legend. (b)~Electric potential $\Psi(x,z)$ and director field $\theta(x,z)$ at final time $t =50/N$ with director configuration shown in red.}
    \label{fig:A=50_h0=0.5}
\end{figure}

 We now discuss the effect of anchoring strength on a thinner film. Figures~\ref{fig:A=1_h0=0.5} and~\ref{fig:A=50_h0=0.5} illustrate the evolution for a film of initial height $h_0 = 0.5$, with weak homeotropic free surface anchoring of strengths $\scA = 1$ and $\scA = 50$ respectively, together with the corresponding electric potential and director field at the final computed time. The corresponding evolution with $\scA=10$ is shown in Fig.~\ref{h:h=0.5_A=10}. As in that figure, we observe here that the film surface exhibits wrinkle formation and develops a distinctly non-sinusoidal profile. The formation of wrinkling patterns appears to depend primarily on the initial film height and not on free surface anchoring strength. We expand on this conjecture in the next section when we discuss how planar anchoring conditions on both boundaries affect the surface height evolution.  
 
 There are, however, some qualitative differences in the evolution as anchoring strength varies.
 First, for weaker anchoring strength and in the regions between electrodes where the electric potential is near zero, the film height slowly increases in time (see Fig.~\ref{fig:h=0p5_A=1} at $x=1$ and $x = 3$). By contrast, as anchoring strength increases to $\scA = 10$ and $\scA = 50$, at the same locations the film height decreases (see Fig.~\ref{h:h=0.5_A=10}(a) and Fig.~\ref{fig:h=0.5_A=50}). Moreover, the director orientation at the free surface deviates significantly from the preferred orientation when $\scA = 1$; in fact for this weakest anchoring strength the director orientation is planar throughout the film in the regions between electrodes, where the electric field is weakest: $\theta = \frac{\pi}{2}$ is a stable (constant) steady solution to the governing equations~(cf. Ref.~\cite{Mema2017}). The strong planar substrate anchoring dominates over both the weak homeotropic anchoring and the weak electric field in regions between the electrodes, forcing the director to align parallel to the substrate throughout the layer: $\theta(1,z) = \theta(3,z) = \pi/2$. 
 
 As the anchoring strength increases to $\scA = 10$ and $\scA = 50$ the constant stable steady state ceases to exist, even in the weak-field regions; the director bends across the layer at all $x$-locations, deviating only slightly from the preferred homeotropic orientation at $z=h$.  In regions above the electrodes (assuming positive dielectric anisotropy), the electric field helps the director align closely with the free surface homeotropic anchoring orientation, independently of the anchoring strength.

\begin{figure}[ht]
  	{\center
\hspace*{-0.2in}
    \subfigure[]{
    \includegraphics[width=0.5\textwidth]{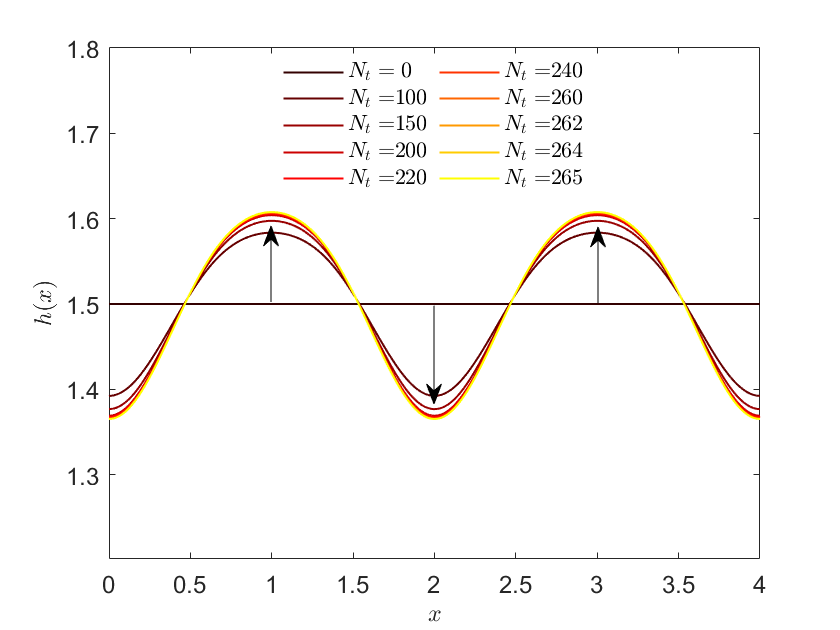}}}
    \subfigure[]{
    \includegraphics[width=0.5\textwidth]{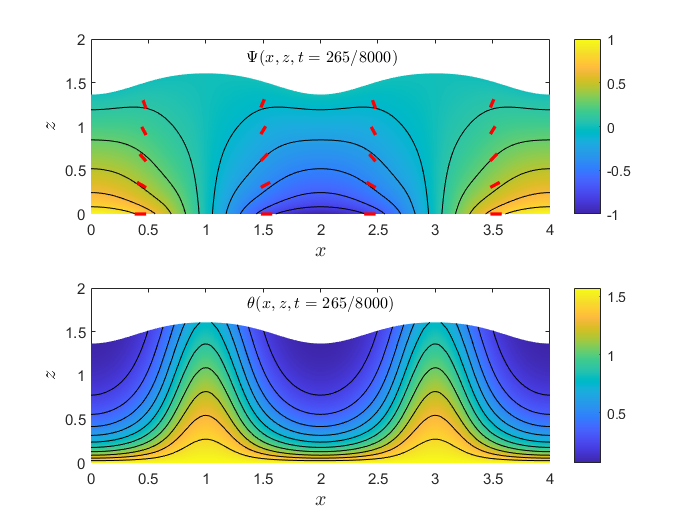}}
    \caption{Evolution of an initially flat film, $h_0 =1.5$ with weak homeotropic free surface anchoring of dimensionless strength $\scA = 1$ and parameters: $\mathscr{C}=\mathscr{N}=0.625$ and $\scD = 10$.  Various times $t = N_t/N$ are shown in (a) where $N = 8000$ and $N_t$ is specified in the legend. (b)~Electric potential $\Psi(x,z)$ at final time $t =265/N$ with director configuration (shown in red) at that time.}
    \label{fig:A=1_h0=1.5}
\end{figure}

\begin{figure}[ht]
	{\center
\hspace*{-0.2in}
    \subfigure[]{
    \includegraphics[width=0.5\textwidth]{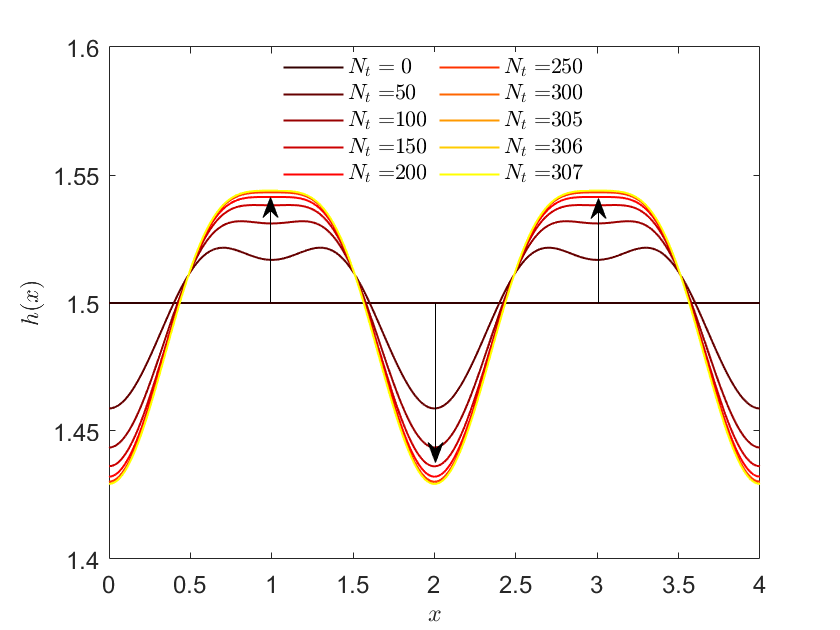}}}
    \subfigure[]{
    \includegraphics[width=0.5\textwidth]{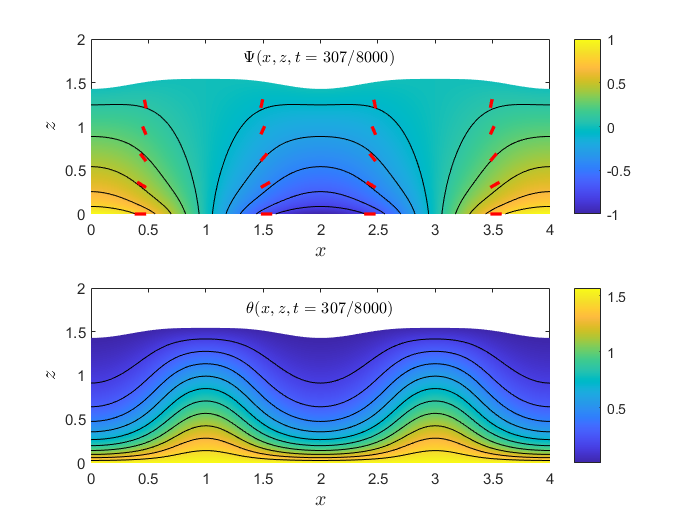}}
    \caption{Evolution of an initially flat film, $h_0 =1.5$ with weak homeotropic free surface anchoring of dimensionless strength $\scA = 50$ and parameters: $\mathscr{C}=\mathscr{N}=0.625$ and $\scD = 10$. Various times $t = N_t/N$ are shown in (a) where $N = 8000$ and $N_t$ is specified in the legend. (b)~Electric potential $\Psi(x,z)$ at final time $t = 307/N$ with director configuration (shown in red) at that time.}
    \label{fig:A=50_h0=1.5}
\end{figure}

 Finally, we comment briefly on the effect of anchoring strength for thicker films:
 Figures~\ref{fig:A=1_h0=1.5} and~\ref{fig:A=50_h0=1.5} illustrate the evolution of an initially flat film of thickness $h_0=1.5$ with strong planar substrate anchoring and weak homeotropic free surface anchoring, of strengths $\scA = 1$ and $\scA = 50$ (respectively). The analogous evolution with $\scA=10$ is shown in Fig.~\ref{h:h=1.5_A=10}(a): similar to that case, both films reach a steady-state near-sinusoidal profile (shown in Figs~\ref{fig:A=1_h0=1.5}(a) and~\ref{fig:A=50_h0=1.5}(a)). Small differences in the free surface height are observed for the different cases: as anchoring strength increases, the amplitude of the perturbations to the steady-state surface height decreases, and the film height maxima become more diffuse. In all cases the maxima are achieved at the points mid-way between the electrodes, $x = 1$ and $x = 3$. For the smallest anchoring strength $\scA=1$ the maximum film height at the final time $t_{\rm f}$, $h(x,t_{\rm f}) \approx 1.6$. As anchoring strength increases to $\scA = 10$ and $\scA = 50$, the maximum film height decreases slightly: $h(x,t_{\rm f}) \approx 1.55$. We note also that the director at the free surface deviates slightly from the preferred homeotropic orientation at the weakest anchoring strength $\scA = 1$ when compared to the simulations with stronger anchoring strengths $\scA = 10$ and $\scA = 50$. As expected, however, this difference is less pronounced than for thinner films.  

\subsubsection{Planar Anchoring at Both Boundaries \& Negative Dielectric Anisotropy}

Motivated by the wish to distinguish between the effects of internal elasticity due to director field distortions, and dielectric effects due solely to the electric field, we now discuss the evolution of an initially flat NLC film with {\it negative} dielectric anisotropy, subject to strong planar anchoring at the lower substrate and weak planar anchoring of strength $\scA = 10$ at the free surface. With dielectric anisotropy parameter $\scD<0$ the nematic molecules align perpendicular to the electric field, rather than parallel to it, so that all external effects here favor a director field that orients parallel to the substrate, with $\theta =\pi/2$ throughout and zero bulk distortion (the elastic contribution $W_{\rm e}=\theta_z^2/2$ to the bulk free energy density in Eq.~(\ref{W_nondim}) is zero). We present results for three different initial film thicknesses, $h_0=0.3,1.0,1.5$ and compare them with the analogous simulations presented above (with strong planar anchoring at the substrate and weak homeotropic anchoring at the free surface, which we refer to as the planar-homeotropic case). Here, we choose the smaller value $h_0 = 0.3$ as representative of a thin film (rather than $h_0 = 0.5$ considered earlier) in order to observe free-surface wrinkling patterns, which are not exhibited when $h_0 = 0.5$.

\begin{figure}[h!]
	{\center
\hspace*{-0.2in}
    \subfigure[]{\includegraphics[width=0.5\textwidth]{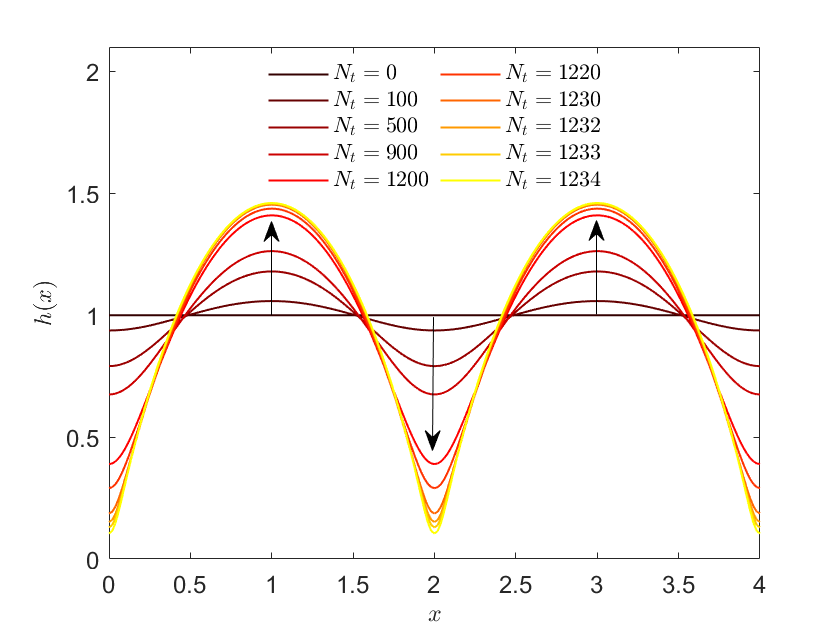}}}
    \subfigure[]{\includegraphics[width=0.5\textwidth]{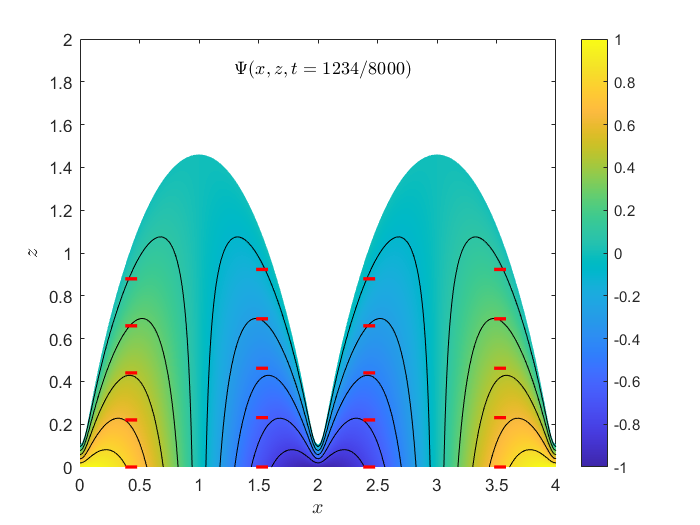}}
    \caption{Evolution of initially flat film, $h_0=1.0$ with strong planar substrate anchoring and weak planar free surface anchoring of dimensionless strength $\scA= 10$ and parameters $\mathscr{C} = \mathscr{N}=0.625$, $\scD = -10$.  Various times $t = N_t/N$ are shown in (a) where $N = 8000$ and $N_t$ is specified in the legend. (b)~Electric potential $\Psi(x,z)$ at final time $t = 1234/N$ with director configuration shown in red.}
    \label{fig:h=1.0_A=10_planar}
\end{figure}

Figure~\ref{fig:h=1.0_A=10_planar} shows the evolution of an initially flat film of intermediate height $h_0 = 1.0$. As noted above, the director aligns parallel to the substrate throughout the layer (see Fig.~\ref{fig:h=1.0_A=10_planar}(b)), which illustrates the director configuration superimposed on the electric field potential at the final computed time. There are similarities to the evolution of the initially flat planar-homeotropic film of the same height shown in Fig.~\ref{fig_h:h=1.0_A=10}(a): the film thickens around the points $x = 1$ and $x = 3$ mid-way between the electrodes where electric field gradients are small, and begins to thin directly above the electrode midpoints at $x = 0,2$ and $4$, where electric field gradients are large. There are some differences, however: the surface height profile in Fig.~\ref{fig:h=1.0_A=10_planar} has well-defined peaks exactly at $x = 1$ and $x = 3$; while the surface height profile for the planar-homeotropic case in Fig.~\ref{fig_h:h=1.0_A=10} has more diffuse maxima. We attribute the diffuseness to the competing forces that the director experiences in the planar-homeotropic case. This competition becomes more intense as free surface anchoring strength is increased, hence the film maxima are even more diffuse at the largest anchoring strength considered ($\scA=50$ in Fig.~\ref{fig:A=50_h0=1.0}(a)). 

\begin{figure}[h!]
	{\center
\hspace*{-0.2in}
    \subfigure[]{\includegraphics[width=0.5\textwidth]{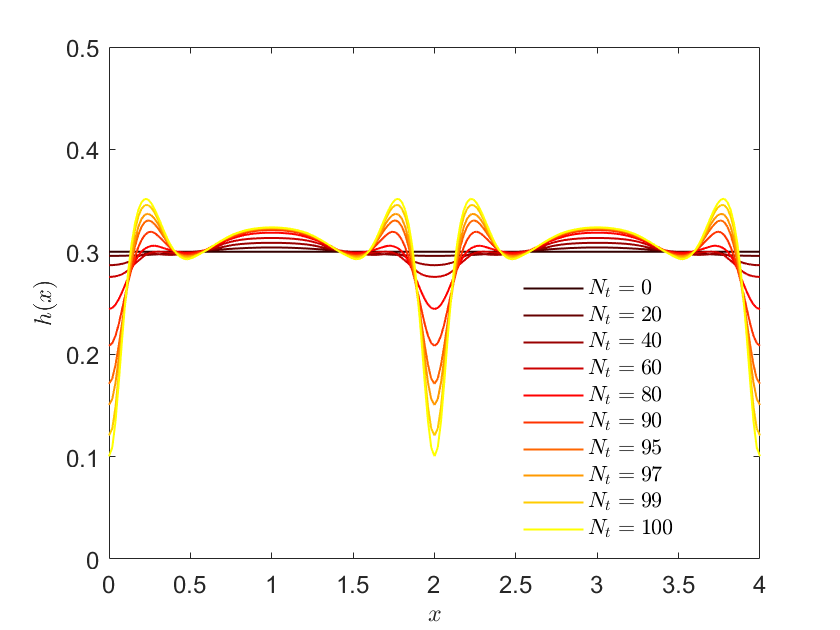}}}
    \subfigure[]{\includegraphics[width=0.5\textwidth]{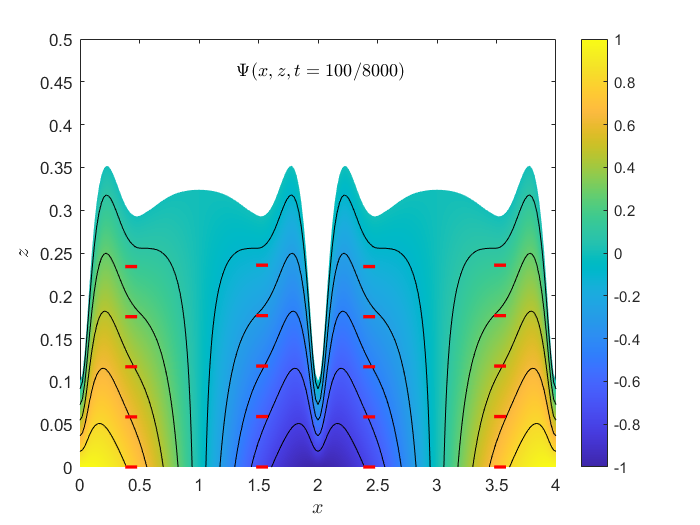}}
    \caption{Evolution of an initially flat film, $h_0 = 0.3$ with strong planar substrate anchoring and weak planar free surface anchoring of dimensionless strength $\scA = 10$ and parameters: $\mathscr{C}=\mathscr{N} = 0.625$ and $\scD = -10$. Various times $t = N_t/N$ are shown in (a) where $N = 8000$ and $N_t$ is specified in the legend. (b)~Electric potential $\Psi(x,z)$ at final time $t = 100/N$ with director configuration shown in red.}
    \label{fig:h=0.3_A=10_planar}
\end{figure}

We next discuss the thin film of initial height $h_0 = 0.3$. The evolution, shown in Fig.~\ref{fig:h=0.3_A=10_planar}, is compared to that of the flat planar-homeotropic film of height $h_0 = 0.5$ as shown in Fig.~\ref{h:h=0.5_A=10}(a) (see also Figs.~\ref{fig:h=0p5_A=1} and~\ref{fig:h=0.5_A=50} for different free surface anchoring strengths). Complex, strongly non-sinusoidal wrinkling patterns emerge in all cases, with the films thinning/thickening in regions of large/small electric field gradients. We here highlight key differences between the cases: first, note that in Fig.~\ref{fig:h=0.3_A=10_planar} the fluid collects in regions mid-way between the electrodes where the electric potential gradient is small, forming local film maxima at $x=1$ and $x = 3$ (local {\it minima} are formed at these points for the planar-homeotropic case in Figs.~\ref{h:h=0.5_A=10}, \ref{fig:h=0p5_A=1} and~\ref{fig:h=0.5_A=50}). Second, we observe four small additional peaks in regions of high electric potential gradients, near $x = 0,$ $x=2$ (one on either side) and $x= 4$ that were not observed in the earlier simulations  (where surface height {\it decreases} over time at these locations). Despite these differences, the re-emergence of wrinkling patterns for these different anchoring conditions reinforces our conjecture that they are ubiquitous for sufficiently thin films (though the details of the anchoring and dielectric anisotropy may modulate the specific wrinkling patterns observed).

\begin{figure}[h!]
	{\center
\hspace*{-0.2in}
    \subfigure[]{
    \includegraphics[width=0.5\textwidth]{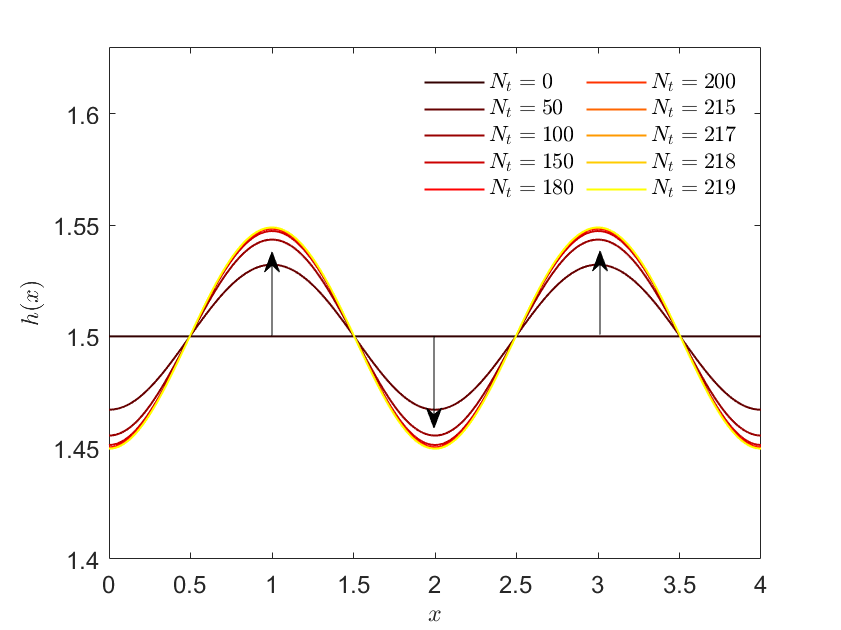}}}
    \subfigure[]{
    \includegraphics[width=0.5\textwidth]{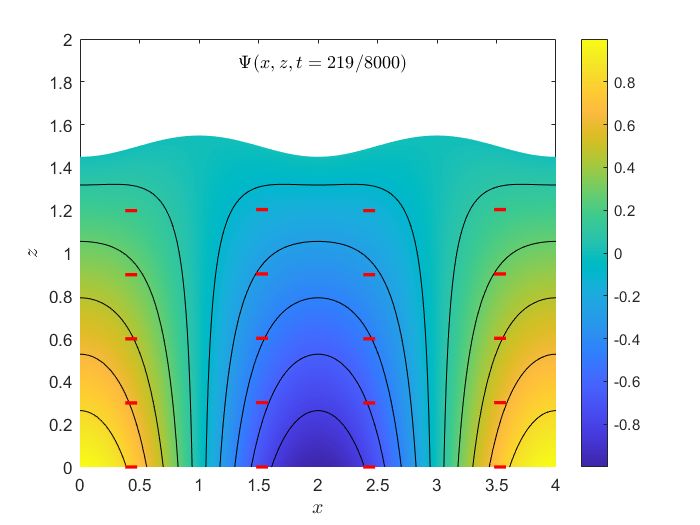}}
    \caption{Evolution of an initially flat film, $h_0 =1.5$ with strong planar substrate anchoring and weak planar free surface anchoring of dimensionless strength $\scA = 10$ and parameters: $\mathscr{C}=\mathscr{N}=0.625$ and $\scD = -10$. Various times $t = N_t/N$ are shown in (a) where $N = 8000$ and $N_t$ is specified in the legend. (b)~Electric potential $\Psi(x,z)$ at final time $t = 219/N$ with director configuration shown in red.}
    \label{fig:h=1.5_A=10_planar}
\end{figure}

Finally, we compare the evolution of a thick film, $h_0=1.5$, to that observed in the planar-homeotropic case shown previously in Fig.~\ref{h:h=1.5_A=10} (see also Figs.~\ref{fig:A=1_h0=1.5} and~\ref{fig:A=50_h0=1.5} for different free surface anchoring strengths).  Figure~\ref{fig:h=1.5_A=10_planar} shows the evolution of a film with strong planar substrate anchoring, weak planar free surface anchoring of strength $\scA=10$, and negative dielectric anisotropy. As in the planar-homeotropic cases of Figs.~\ref{h:h=1.5_A=10},~\ref{fig:A=1_h0=1.5} and~\ref{fig:A=50_h0=1.5} we observe that the film thickens in regions between electrodes where electric field gradients are small and thins directly above electrodes, where electric field gradients are large. In addition, and again similar to the  planar-homeotropic cases, a steady state is quickly reached. The surface height profile in Fig.~\ref{fig:h=1.5_A=10_planar} has well-defined maxima at $x = 1$ and $x=3$ while the corresponding planar-homeotropic surface height profile in Fig.~\ref{h:h=1.5_A=10} exhibits more diffuse maxima at these points. As with the intermediate film thickness $h_0=1$ simulations, we conjecture that this difference is due to the competition between different forces on the NLC molecules in the planar-homeotropic case.

Finally, we note that from a modeling perspective we anticipated that this last-considered case of NLC film evolution with planar anchoring conditions and negative dielectric anisotropy should provide results most analogous to those for an isotropic dielectric liquid, due to the uniform director orientation throughout the layer. However, when comparing our results with the experimental observations reported by Brown {\it et al.}~\cite{Brown2009}, we see that a closer qualitative match is obtained with the earlier planar-homeotropic simulations of Figs.~\ref{fig_h:h=1.0_A=10}--\ref{fig:A=50_h0=1.5} than with the results of this section. Since a NLC, even with uniform director orientation, is not the same as an isotropic dielectric, we conclude that other factors must account for the observed differences in evolution. We believe that the stronger qualitative similarities between the experiments and our simulations for the planar-homeotropic case are at least partly coincidental.

\section{Conclusion}

We have presented a mathematical model that describes the flow of a thin NLC film in the presence of a nonuniform electric field. Specifically, we consider a thin layer of NLC coated on a substrate $z=0$ that contains embedded planar interlaced electrodes, the effect of which we approximate by a periodic electric field profile on $z=0$ in the $x$-coordinate direction. The mathematical model derived consists of a 4th order nonlinear parabolic partial differential equation for the height $h(x,t)$ of the film, coupled to a boundary value system for the electric potential $\Psi(x,z,t)$ and the director field $\mathbf{n}=(\sin\theta,\cos\theta)$, characterized in our 2D setup by a single angle $\theta(x,z,t)$.

Numerical techniques were used to investigate the temporal evolution of an initially flat film, with initial dimensionless height representative of thick ($h_0 = 1.5$), intermediate ($h_0 = 1.0$) or thin ($h_0 = 0.5$ or $h_0 = 0.3$) films. For thick films, the surface height evolves to an undulating steady-state profile in all cases considered: fluid collects in the regions between electrodes where the electric potential is near zero and electric field gradients are small, and thins at the electrode midpoints, where the electric and director field gradients are large.  For thin and intermediate thickness films, fluid accumulates in regions of high electric potential and director field gradients, migrating from regions of small gradients. Unlike the thicker films, thin and intermediate films thin significantly in these high-gradient regions; no steady state is found and our simulations are halted when accuracy is lost in the solution for the director field. For the thinnest films, complex wrinkling patterns form, with the film thickening at the edges of the electrodes and further thinning observed at the electrode midpoints where the electric potential gradients are large, leading to global film minima at these points.  Between the electrodes, film thinning is suppressed and local minima form above $x = 1$ and $x = 3$. 

Additionally, we investigated the effect of different anchoring conditions on NLC film evolution; specifically we  considered the effect of free surface homeotropic anchoring strength $\scA$  on the surface height evolution by considering three different values:  $\scA =1$, $\scA = 10$ and $\scA = 50$. Changing $\scA$ primarily affects the director configuration and its gradients, which in turn affect the surface height evolution. We observed that as $\scA$ is increased the maxima of the film surface profile become more diffuse, possibly due to increased competition of different forces within the film.  For sufficiently weak anchoring, in the thinnest film simulated ($h_0=0.5$) we also observe regions in the film where the director field is uniform and planar (Fig.~\ref{fig:A=1_h0=0.5}), reflecting the fact that $\theta=\pi/2$ is a stable steady solution to the governing equations under the local film conditions.

Finally, with the goal of gaining insight into the behavior of film evolution when there are no elastic forces within the film layer, we considered an NLC film with negative dielectric anisotropy, subject to strong planar anchoring at the lower substrate and weak planar free surface anchoring. Negative dielectric anisotropy means that the nematic molecules align perpendicular to the direction of the electric field instead of parallel to it; the director in this case is uniform throughout the layer, and the elastic energy is negligible. While surface height evolution was generally similar to that of analogous films with planar-homeotropic anchoring, the films with planar anchoring and negative dielectric anisotropy had well-defined maxima, which we attribute to the lack of competition between opposing forces in the film. The thinnest films in this case exhibit complex non-sinusoidal wrinkling profiles, behavior that appears to depend primarily on the initial film height. 

We note that our numerical results are in qualitative agreement with the experiments of Brown {\it et al.}~\cite{Brown2010} who investigate dielectrophoresis of an isotropic dielectric fluid over interlaced electrodes. They observe that fluid collects above electrodes, where electric potential gradients are highest, and is removed from regions between electrodes where electric potential gradients are small. A similar experiment carried out by the same group~\cite{Brown2009} demonstrated that as the initial height of the dielectric film decreases, wrinkled, strongly non-sinusoidal profiles emerge, similar to those in our thinnest film simulations. Though we anticipated that the NLC film with planar free surface anchoring and negative dielectric anisotropy might be more analogous to the isotropic dielectric liquid (having zero net elastic energy), this expectation was not borne out: our simulations of planar-homeotropic flat films provide the best agreement with the experimental results of Brown {\it et al.}~\cite{Brown2010}. We hope that our new model and computational results will inspire future experimental investigations that may provide further insight and lead to improvement of our mathematical model. 

\section{Acknowledgments}
This work was supported by the NSF under Grant No. DMS 1615719.

\bibliographystyle{abbrv}
\bibliography{bibliography}

\end{document}